\documentclass[aps,pra,reprint,superscriptaddress,nopacs,nofootinbib,longbibliography]{revtex4-1}

\usepackage{amsmath,amssymb,latexsym}
\usepackage{graphicx}
\graphicspath{{../fig/}}
\usepackage{hyperref}
\usepackage{mathrsfs}
\usepackage{amscd}
\usepackage{color}
\usepackage{comment}
\DeclareSymbolFont{symbols}{OMS}{cmsy}{m}{n}
\DeclareSymbolFont{largesymbols}{OMX}{cmex}{m}{n}
\renewcommand{\bm}[1]{\boldsymbol #1}

\begin{document}

\title{
Two-dimensional coherent spectroscopy of disordered superconductors in the narrow-band and broad-band limits
}

\author{Naoto Tsuji}
\affiliation{Department of Physics, University of Tokyo, Hongo, Tokyo 113-0033, Japan}
\affiliation{RIKEN Center for Emergent Matter Science (CEMS), Wako 351-0198, Japan}

\begin{abstract}
We theoretically analyze two-dimensional coherent spectroscopy (2DCS) signals for disordered superconductors
in two limits: 
One is the narrow-band limit with sinusoidal pulse waves, and the other is the broad-band limit with delta-function pulses.
While the 2DCS signal in the narrow-band limit
is related to the third-order nonlinear susceptibilities $\chi^{(3)}(3\Omega; \Omega, \Omega, \Omega)$ 
(third harmonic generation)
and $\chi^{(3)}(\Omega; \Omega, \Omega, -\Omega)$ (ac Kerr effect),
we find that in the broad-band limit the signal along the diagonal and horizontal lines in the two-dimensional frequency space
is related to another nonlinear susceptibility $\chi^{(3)}(\Omega; \Omega, 0, 0)$ (dc Kerr effect).
We numerically evaluate those susceptibilities for a lattice model of superconductors
based on the BCS mean-field theory and self-consistent Born approximation
for impurities. The 2DCS signals in the narrow-band and broad-band limits 
show threshold and resonance behaviors at the superconducting-gap frequency, respectively,
whose physical origin is discussed in light of quasiparticle and Higgs-mode excitations.
\end{abstract}


\date{\today}

\maketitle

\section{Introduction}
\label{sec: introduction}

Two-dimensional coherent spectroscopy (2DCS) \cite{Mukamel1995, Cho2008, Hamm2011, Biswas2022, Fresch2023, Liu2025, Huang2025} is recently adopted as a tool to probe nonlinear optical response of
various quantum materials using multiple laser pulses. By controlling delay times between pulses,
one can map out the signal in a multi-dimensional frequency space, which brings fruitful information
that is hard to access with other spectroscopy techniques such as absorption and reflection spectroscopies.
In particular, one can subtract linear response components systematically by taking the difference between different pulse configurations, 
which allows one to extract nonlinear response components
accurately even at the first harmonic frequency of injected pulses.

The 2DCS is useful in detecting collective modes in a condensed phase of materials including quantum magnets and superconductors
\cite{Tsuji2024}.
The presence of those collective modes tells us key aspects of quantum phases, e.g., symmetries, fluctuations, phase transitions, hidden orders, and so on.
In superconductors, there exists the Higgs mode \cite{Anderson1958, Schmid1968, PekkerVarma2015, ShimanoTsuji2020}, 
i.e., an amplitude mode of the superconducting order parameter, which does not
linearly couple to electromagnetic fields in usual situations (an exception is the case where Cooper pairs have nonzero center-of-mass momentum \cite{Moor2017, Nakamura2019, Kubo2024, Wang2025, Nagashima2025, Niederhoff2025}).
Previously, the Higgs mode has been observed by a nonlinear response setup such as Raman scattering \cite{SooryakumarKlein1980, Sooryakumar1981, LittlewoodVarma1981, LittlewoodVarma1982, Measson2014, Grasset2018, Grasset2019}, 
pump-probe spectroscopy \cite{Matsunaga2013, Matsunaga2014, Katsumi2018, Katsumi2020},
and third harmonic generation \cite{Matsunaga2014, Tsuji2015, Matsunaga2017} (see also Refs.~\cite{Chu2020, Kovalev2021, Isoyama2021, Wang2022, Katsumi2023, Kim2024}). 
On the other hand, it is not straightforward to extract the contribution of the Higgs mode separately,
as individual excitations of quasiparticles also contribute to the signal at similar frequencies (which correspond to the superconducting gap size).
This raises a question of finding other measurement protocols that can detect the effect of collective modes selectively.

Recently, the 2DCS has been applied to study nonlinear response of NbN and MgB$_2$ superconductors in the terahertz regime \cite{Katsumi2024, Katsumi2025}.
For NbN, the temperature dependence of the first harmonic contribution of the 2DCS signal with a narrow-band pulse 
shows a peak when the frequency agrees with the superconducting gap ($\Omega=2\Delta$).
In contrast, the 2DCS for MgB$_2$ superconductors does not show such a peak but exhibits monotonic temperature dependence. 
The interpretation of the results has been discussed in connection to collective modes of superconductors.
The 2DCS has also been employed to study nickelate superconductors \cite{Cheng2025} and cuprates \cite{Chaudhuri2025}.
Theoretically, the 2DCS has been analyzed for multiband superconductors \cite{Mootz2024}, layered superconductors \cite{GomezSalvador2024},
correlated insulators \cite{ChenWerner2025}, and antiferromagnets \cite{Chen2025}.

Motivated by the recent experiments, in this paper we study the 2DCS for disordered superconductors in two limits: 
One is the narrow-band limit where pulses are modeled as sinusoidal waves, and the other is the broad-band limit
where pulses are treated as delta functions. While the first harmonic component of the 2DCS signal is described by
the ac Kerr susceptibility $\chi^{(3)}(\Omega; \Omega, \Omega, -\Omega)$,
we find that the 2DCS signal in the broad-band limit is related to the dc Kerr susceptibility $\chi^{(3)}(\Omega; \Omega, 0, 0)$.
We also derive a general formula for the 2DCS with general pulse wave forms, which is given by the frequency integral
of the third-order nonlinear susceptibility with different frequency combinations. 
We numerically evaluate the ac and dc Kerr susceptibilities for a lattice model of disordered superconductors.

As suggested for the optical conductivity \cite{MattisBardeen1958}
and third harmonic generation \cite{Jujo2018, MurotaniShimano2019, Silaev2019, TsujiNomura2020, Seibold2021, Derendorf2024}, disorders play a pivotal role in understanding
(non)linear response of superconductors (phonon retardation effects also play a role \cite{Tsuji2016}). 
In fact, the Higgs mode has been shown to contribute dominantly to the THG resonance 
at half of the gap frequency ($\Omega=\Delta$) in the dirty regime, while the Higgs mode becomes less dominant than quasiparticles
in the clean regime \cite{Cea2016}. To treat disorders in the dirty regime, the system has to satisfy the following length scale separation:
\begin{align}
\frac{v_F}{W}
&\ll
\frac{v_F}{\gamma}
\ll
\frac{v_F}{2\Delta}
\ll
L.
\label{scale separation}
\end{align}
Here $v_F$ is the Fermi velocity, $W$ is the bandwidth of electrons, $\gamma$ is the disorder scattering rate,
$2\Delta$ is the superconducting gap, and $L$ is the linear system size. Note that we set $\hbar=1$ throughout this paper.
In the dirty regime, the coherence length $\frac{v_F}{2\Delta}$ must be longer than the mean free path $\frac{v_F}{\gamma}$.
To avoid finite-size effects, the linear system size should be taken to be much longer than the coherence length.
When $\gamma$ becomes comparable to $W$, the effect of localization begins to emerge, which is far from experimental situations.
Typically, one has $L\sim 10^{-2}$ [m], $v_F\sim 10^6$ [m/s], $2\Delta \sim 10^{12}$ [s$^{-1}$], 
$\frac{v_F}{2\Delta}\sim 10^{-6}$ [m], $\frac{v_F}{\gamma}\sim 10^{-8}$-$10^{-7}$ [m], and $\frac{v_F}{W}\sim 10^{-9}$ [m]
for dirty superconductors in experiments, satisfying the relation (\ref{scale separation}).

The previous analysis on the 2DCS for disordered superconductors has focused on the narrow-band limit of pulses,
and has been based on the simulation of the mean field theory with random disorders in real space \cite{Katsumi2024}.
While the effect of disorders is treated exactly in this approach, it is not easy to satisfy the clear scale separation (\ref{scale separation}),
for which one typically requires $L\gtrsim 100a$ in a lattice model ($a$ is the lattice constant) \cite{remark}. 
In two dimensions, for instance, the number of lattice sites reaches $100\times 100$, which is computationally demanding.
Furthermore, the finite-size scaling is sometimes challenging: For example, there appears a logarithmic correction in the system size dependence of conductance
in two dimensions \cite{Abrahams1979, LeeFisher1981, Rammer}.

To circumvent those problems, we employ the self-consistent Born approximation \cite{AbrikosovGorkov1959, TsujiNomura2020}, 
which is a semiclassical treatment of disorders that allows
one to approach the desired scale separation (\ref{scale separation}). In fact, we can take $100\times 100$ lattice sites within reasonable numerical cost. 
Since the self-consistent Born approximation is a kind of the mean-field theory for disorders, the results do not significantly depend on
the spatial dimension of the system (describing the behavior in large dimensions), 
and the system size dependence can be numerically well controlled.
Practically, the self-consistent Born approximation should be valid in experimentally realistic situations with $\gamma\ll W$.

Using the self-consistent Born approximation for disorders and BCS approximation for the pairing interaction,
we obtain the ac and dc Kerr susceptibilities for disordered superconductors. The results show that the ac Kerr susceptibility
shows threshold behavior, i.e., the intensity starts to grow above the superconducting gap frequency ($\Omega\ge 2\Delta$), in both the clean and dirty regimes.
Physically, this behavior is mainly mediated by the zero-frequency Higgs mode which is far off-resonant.
On the other hand, the dc Kerr susceptibility exhibits a resonance peak at the gap frequency ($\Omega=2\Delta$),
which is mostly mediated by the Higgs mode carrying the frequency $\Omega$ in the dirty regime.
We examine the temperature and disorder dependence of the ac and dc Kerr susceptibilities, and discuss
the relation to the experiments.

The paper is organized as follows. In Sec.~\ref{sec: linear response}, we review the formalism of the linear optical response
in the narrow-band and broad-band limits as a warm up. In Sec.~\ref{sec: 2DCS general},
we formulate the 2DCS with general pulse wave forms, and derive a formula that relates the 2DCS with the nonlinear susceptibilities.
In Sec.~\ref{sec: 2DCS narrow-band} and Sec.~\ref{sec: 2DCS broad-band}, we treat the narrow-band and broad-band limits of the 2DCS, respectively.
In Sec.~\ref{sec: numerical}, we demonstrate the numerical results on the ac and dc Kerr susceptibilities for a lattice model of disordered superconductors.
The paper is summarized in Sec.~\ref{sec: summary} with the outlook of future directions of the study.

\section{Linear optical response in the narrow-band and broad-band limits}
\label{sec: linear response}

In this section, we overview the derivation of the linear optical response functions
in the narrow-band and broad-band limits, which can be extended to nonlinear responses including the 2DCS in later sections.
It has been known that the two limits are intimately related to each other, and both of them
can be described by the same response function.

In the narrow-band limit, an injected pulse is a monochromatic wave
defined by a vector potential
\begin{align}
\bm A(t)
&=
\bm A e^{-i\Omega t}
\end{align}
with a constant amplitude vector $\bm A$.
The measured current expanded up to the first order in the drive field is written as
\begin{align}
\bm j(t)
&=
\sigma(\Omega)\bm E(t),
\end{align}
where $\sigma(\Omega)$ is the optical conductivity (i.e., the linear response coefficient), 
and $\bm E(t)=-\partial_t \bm A(t)$ is the driving electric field.
Throughout the paper, we assume that applied pulse fields and induced currents are polarized along
the same direction for simplicity, and suppress the vector notation below (for example, $j(t)$ should be understood
as $\bm e\cdot \bm j(t)$ where $\bm e$ is the polarization unit vector).

According to the standard linear response theory, the optical conductivity is given by
\begin{align}
\sigma(\Omega)
&=
\frac{1}{i\Omega}\chi^{(1)}(\Omega),
\label{sigma}
\end{align}
where $\chi^{(1)}(\Omega)$ is formally defined as
\begin{align}
\chi^{(1)}(\Omega)
&=
\int_{-\infty}^\infty d\bar t\, e^{i\Omega (t-\bar t)}\frac{\delta j(t)}{\delta A(\bar t)}\bigg|_{A=0}.
\label{chi^(1)}
\end{align}
Note that the right hand side of Eq.~(\ref{chi^(1)}) does not depend on $t$ due to the time-translation symmetry.
The current is directly written in terms of $\chi^{(1)}$ as
\begin{align}
j(t)
&=
\chi^{(1)}(\Omega) A(t)
\end{align}
up to the first order in $A$.


In the broad-band limit, the driving field is modeled with a delta-function pulse,
\begin{align}
E(t)
&=
A\delta(t).
\end{align}
Here $A$ is the amplitude of the pulse having the dimension of the vector potential.
We fix the time at which the delta-function pulse is applied at $t=0$.
The corresponding vector potential is given by
\begin{align}
A(t)
&=
-A\theta(t).
\label{A theta}
\end{align}
One can expand the current measured at time $t$ up to the first order with respect to $A$,
\begin{align}
j(t)
&=
\int_{-\infty}^\infty d\bar t\, \frac{\delta j(t)}{\delta A(\bar t)}\bigg|_{A=0} A(\bar t)
\label{current broad band}
\end{align}
We define the Fourier transform of $j(t)$ as
\begin{align}
j(\Omega)
&=
\int_{-\infty}^\infty dt\, e^{i\Omega t}j(t).
\end{align}
Substituting Eq.~(\ref{A theta}) into Eq.~(\ref{current broad band}) and using the relation
($\eta$ is a positive infinitesimal constant),
\begin{align}
\theta(t)
&=
i\int_{-\infty}^\infty \frac{d\omega}{2\pi} \frac{e^{-i\omega t}}{\omega+i\eta},
\label{step function}
\end{align}
we obtain
\begin{align}
j(\Omega)
&=
-iA\int_{-\infty}^\infty \frac{d\omega}{2\pi} 
\int_{-\infty}^\infty dt\, e^{i(\Omega-\omega) t}
\frac{\chi^{(1)}(\omega)}{\omega+i\eta}
\notag
\\
&=
-iA\frac{\chi^{(1)}(\Omega)}{\Omega+i\eta}
=
A\sigma(\Omega).
\end{align}
Thus, the linear response against the delta-function pulse (i.e., the broad-band limit) is also described by
the optical conductivity spectrum. Note, however, that the meaning of $\Omega$ is different between the two limits:
$\Omega$ is the frequency of the monochromatic wave in the narrow-band limit, while $\Omega$ is the frequency
of the Fourier transformed current in the broad-band limit.

In the following sections, the formalism can be straightforwardly extended to multi-pulse cases
for the 2DCS, which can be generally represented by a complicated integral of higher-order correlation functions.
In the narrow-band and broad-band limits, the expression is greatly simplified, and
the 2DCS signal is directly related to certain nonlinear susceptibilities without the integral.
Contrary to the case of the linear response, we will see that the two limits are described by different types of correlation functions.

\section{Two-dimensional coherent spectroscopy with general pulses}
\label{sec: 2DCS general}

\begin{figure}[t]
\includegraphics[width=8cm]{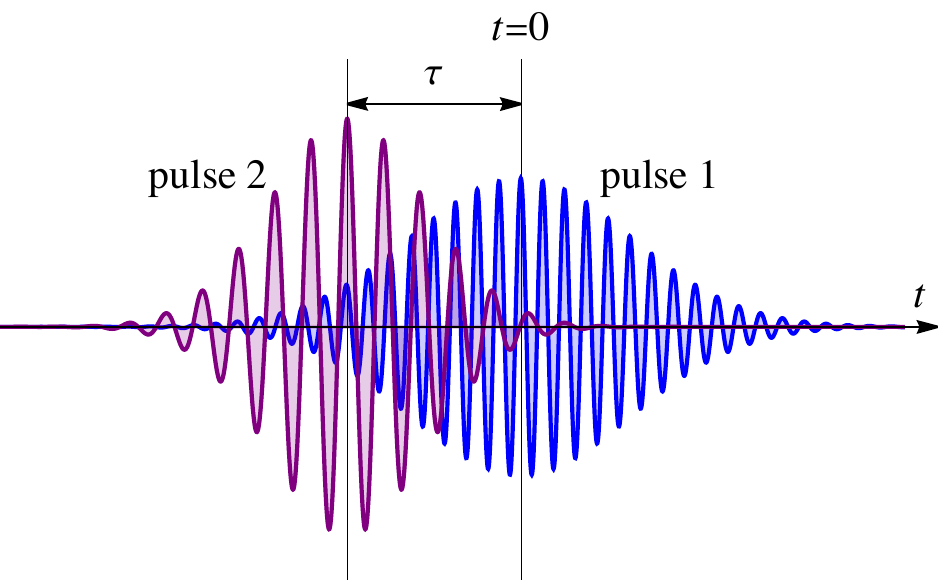}
\caption{Schematic picture of the configuration of two pulses with a time delay $\tau$.}
\label{fig: general pulses}
\end{figure}

Let us consider a general situation where the system is driven by two pulses with arbitrary wave forms.
We call the first pulse injected into the system a `pulse 1', and call the second one a `pulse 2'.
The delay time between the two pulses is denoted by $\tau$. We show the time profile of the two pulses in Fig.~\ref{fig: general pulses}.
The vector potential is written as
\begin{align}
A(t)
&=
A_1(t)+A_2(t+\tau),
\end{align}
where $A_1(t)$ is the wave form of the pulse 1,
and $A_2(t)$ is the wave form of the pulse 2, which is shifted by the delay time $\tau$.

We define the current solely induced by the pulse $i(=1,2)$ as $j_i(t)$.
That is, we turn off the other pulse and measure the current, which is expanded as a function of $A_i(t)$
up to the third order.
Since the second-order term is absent when the system has the inversion symmetry, 
the expansion of $j_i(t)$ reads
\begin{align}
j_i(t)
&=
\int_{-\infty}^\infty d\bar t\, \frac{\delta j(t)}{\delta A(\bar t)}\bigg|_{A=0} A_i(\bar t+\delta_{i,2}\tau)
\notag
\\
&\quad
+\frac{1}{3!} \int_{-\infty}^\infty d\bar t d\bar t' d\bar t'' 
\frac{\delta^3 j(t)}{\delta A(\bar t)\delta A(\bar t')\delta A(\bar t'')}\bigg|_{A=0}
\notag
\\
&\quad\times
A_i(\bar t+\delta_{i,2}\tau)A_i(\bar t'+\delta_{i,2}\tau)A_i(\bar t''+\delta_{i,2}\tau)
\notag
\\
&\quad
(i=1,2).
\end{align}
We also define the total current induced by both the pulse 1 and pulse 2 as $j_{12}(t)$, which is expanded as
\begin{align}
j_{12}(t)
&=
\int_{-\infty}^\infty d\bar t\, \frac{\delta j(t)}{\delta A(\bar t)}\bigg|_{A=0} (A_1(\bar t)+A_2(\bar t+\tau))
\notag
\\
&\quad
+\frac{1}{3!} \int_{-\infty}^\infty d\bar t d\bar t' d\bar t''
\frac{\delta^3 j(t)}{\delta A(\bar t)\delta A(\bar t')\delta A(\bar t'')}\bigg|_{A=0}
\notag
\\
&\quad\times
(A_1(\bar t)+A_2(\bar t+\tau))(A_1(\bar t')+A_2(\bar t'+\tau))
\notag
\\
&\quad\times
(A_1(\bar t'')+A_2(\bar t''+\tau)).
\end{align}

Now the nonlinear current solely proportional to the third power of $A_i$ is defined by subtracting $j_1(t)$ and $j_2(t)$ from $j_{12}(t)$,
\begin{align}
j_{\rm NL}(t, \tau)
&=
j_{12}(t)-j_1(t)-j_2(t),
\label{j_NL}
\end{align}
which can be written as
\begin{align}
j_{\rm NL}(t, \tau)
&=
\frac{1}{2} \int_{-\infty}^\infty d\bar t d\bar t' d\bar t''
\frac{\delta^3 j(t)}{\delta A(\bar t)\delta A(\bar t')\delta A(\bar t'')}\bigg|_{A=0}
\notag
\\
&\quad\times
(A_1(\bar t)A_1(\bar t')A_2(\bar t''+\tau)
\notag
\\
&\quad
+A_1(\bar t)A_2(\bar t'+\tau)A_2(\bar t''+\tau)).
\end{align}
The nonlinear current is split into two parts: One is proportional to $A_1^2A_2$ and the other is proportional to $A_1A_2^2$.
Let us denote these terms as $j_{A_1^2A_2}(t,\tau)$ and $j_{A_1A_2^2}(t,\tau)$, respectively,
which can be distinguished from the intensity dependence of the two pulses.

We first look at the first part, $j_{A_1^2A_2}(t,\tau)$, which can be double Fourier transformed as
\begin{align}
j_{A_1^2A_2}(\omega_t, \omega_\tau)
&=
\int_{-\infty}^\infty dt\, e^{i\omega_t t}\int_{-\infty}^\infty d\tau\, e^{i\omega_\tau \tau} j_{A_1^2A_2}(t, \tau).
\end{align}
By performing the $\tau$ integral, we obtain
\begin{align}
&
j_{A_1^2A_2}(\omega_t, \omega_\tau)
\notag
\\
&=
\frac{1}{2}\int_{-\infty}^\infty dt\, e^{i\omega_t t}
\int_{-\infty}^\infty d\bar t d\bar t' d\bar t'' 
\frac{\delta^3 j(t)}{\delta A(\bar t)\delta A(\bar t')\delta A(\bar t'')}\bigg|_{A=0}
\notag
\\
&\quad\times
\int_{-\infty}^\infty \frac{d\omega}{2\pi}
\int_{-\infty}^\infty \frac{d\omega'}{2\pi}
e^{-i(\omega \bar t+\omega'\bar t'+\omega_\tau \bar t'')}
\notag
\\
&\quad\times
A_1(\omega)A_1(\omega')A_2(\omega_\tau)
\label{j A_1^2A_2 FT}
\end{align}
Now we rewrite the above expression using the third-order current correlation function defined by
\begin{align}
&
\chi^{(3)}(\Omega; \Omega_1, \Omega_2, \Omega_3)
\notag
\\
&=
\frac{1}{3!}
\int_{-\infty}^\infty d\bar t d\bar t' d\bar t'' \, 
e^{i\Omega t-i(\Omega_1\bar t+\Omega_2\bar t'+\Omega_3\bar t'')}
\notag
\\
&\quad\times
\frac{\delta^3 j(t)}{\delta A(\bar t)\delta A(\bar t')\delta A(\bar t'')}\bigg|_{A=0}
\quad
(\Omega=\Omega_1+\Omega_2+\Omega_3).
\label{chi^(3)}
\end{align}
Note that the right-hand side of Eq.~(\ref{chi^(3)}) does not depend on $t$ due to the time-translation symmetry.
The correlation function $\chi^{(3)}(\Omega; \Omega_1, \Omega_2, \Omega_3)$ is invariant
under the exchange of $\Omega_i$ and $\Omega_j$ ($i\neq j$).

Using $\chi^{(3)}$, we can express Eq.~(\ref{j A_1^2A_2 FT}) as
\begin{align}
j_{A_1^2A_2}(\omega_t, \omega_\tau)
&=
3A_2(\omega_\tau)
\int_{-\infty}^\infty \frac{d\omega}{2\pi}
A_1(\omega)A_1(\omega_t-\omega_\tau-\omega)
\notag
\\
&\quad\times
\chi^{(3)}(\omega_t; \omega, \omega_t-\omega_\tau-\omega, \omega_\tau).
\label{j A_1^2A_2 formula}
\end{align}
This is the general formula that we derive for the 2DCS signal proportional to $A_1^2A_2$.
Since $A_2(\omega_\tau)$ is factored out of the integral in Eq.~(\ref{j A_1^2A_2 formula}), 
the signal is enhanced along the horizontal line in the two-dimensional frequency space where $\omega_\tau$
is close to the central frequency of the spectrum $A_2(\omega)$.
A similar formula can be derived for the signal proportional to $A_1A_2^2$,
\begin{align}
j_{A_1A_2^2}(\omega_t, \omega_\tau)
&=
3A_1(\omega_t-\omega_\tau)
\int_{-\infty}^\infty \frac{d\omega}{2\pi}
A_2(\omega)A_2(\omega_\tau-\omega)
\notag
\\
&\quad\times
\chi^{(3)}(\omega_t; \omega, \omega_\tau-\omega, \omega_t-\omega_\tau).
\label{j A_1A_2^2 formula}
\end{align}
In this case, $A_1(\omega_t-\omega_\tau)$ is factored out, meaning that
the signal is enhanced along the line parallel to the diagonal line where $\omega_t-\omega_\tau$ is close to the central frequency
of the spectrum $A_1(\omega)$.

We notice that there is a symmetry between Eq.~(\ref{j A_1^2A_2 formula}) and Eq.~(\ref{j A_1A_2^2 formula}) through the relation,
\begin{align}
j_{A_1^2A_2}(\omega_t, \omega_\tau)
&=
j_{A_2A_1^2}(\omega_t, \omega_t-\omega_\tau).
\end{align}
Thus, $j_{A_1^2A_2}$ and $j_{A_1A_2^2}$ essentially contain the same information.
If the pulse 1 and pulse 2 have the same wave form (i.e., $A_1(t)=A_2(t)$), then the nonlinear current,
$j_{\rm NL}(\omega_t, \omega_\tau)=j_{A_1^2A_2}(\omega_t, \omega_\tau)+j_{A_1A_2^2}(\omega_t, \omega_\tau)$,
has the reflection symmetry,
\begin{align}
j_{\rm NL}(\omega_t, \omega_\tau)
&=
j_{\rm NL}(\omega_t, \omega_t-\omega_\tau).
\end{align}

From Eqs.~(\ref{j A_1^2A_2 formula}) and (\ref{j A_1A_2^2 formula}), we can see that
the 2DCS signal is given by a frequency integral of the third-order nonlinear susceptibility
multiplied by the Fourier spectra of the pulse fields $A_1$ and $A_2$.
By engineering the wave form and choosing an appropriate combination of the frequencies $\omega_t$ and $\omega_\tau$,
one can extract the information of certain part of $\chi^{(3)}(\Omega; \Omega_1, \Omega_2, \Omega_3)$.
Below we study two particular cases, the narrow-band and broad-band limits,
where the 2DCS signal is most directly related to $\chi^{(3)}$ without the frequency integral.

\section{Two-dimensional coherent spectroscopy in the narrow-band limit}
\label{sec: 2DCS narrow-band}

In this section, we focus on the narrow-band limit, that is, the two pulses are both monochromatic waves with infinitely large pulse widths
(we sometimes call those driving fields as pulses, even though the pulse widths are diverging).
In general, the two driving fields can have different frequencies.
Here we limit ourselves to the case in which the two pulses have the same frequency $\Omega$,
which allows us to write down the 2DCS signal in a simple form.
The corresponding vector potential reads
\begin{align}
A(t)
&=
A_1(t)+A_2(t+\tau)
\notag
\\
&=
A_1 \cos(\Omega t)
+A_2 \cos(\Omega (t+\tau)),
\end{align}
where the pulse 1 has an amplitude $A_1$, and the pulse 2 has an amplitude $A_2$ with the delay time $\tau$. 
Since we consider nonlinear responses, we include both $+\Omega$ and $-\Omega$ components
(i.e., $\cos\Omega t=\frac{1}{2}(e^{i\Omega t}+e^{-i\Omega t})$) in the vector potential.
The corresponding Fourier spectra of the pulse fields are given by
\begin{align}
A_1(\omega)
&=
\pi A_1 (\delta(\omega-\Omega)+\delta(\omega+\Omega)),
\\
A_2(\omega)
&=
\pi A_2 (\delta(\omega-\Omega)+\delta(\omega+\Omega)).
\end{align}

\begin{figure}[t]
\includegraphics[width=8cm]{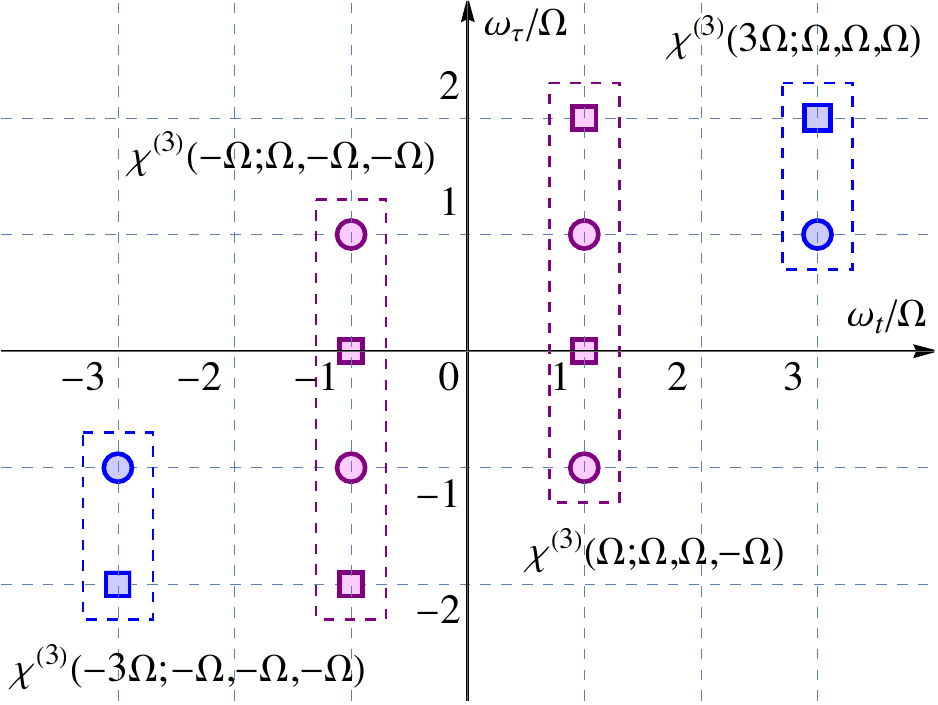}
\caption{2DCS spectrum $j_{\rm NL}(\omega_t, \omega_\tau)$ in the narrow-band limit.
The circles represent the signals proportional to $A_1^2A_2$,
while the squares represent the signals proportional to $A_1A_2^2$. The blue and purple markers correspond to
the third harmonic generation and ac Kerr susceptibilities, respectively.}
\label{fig: 2d spectrum narrow-band}
\end{figure}

Using the formula (\ref{j A_1^2A_2 formula}), we obtain the 2DCS signal proportional to $A_1^2A_2$ in the narrow-band limit,
\begin{align}
&
j_{A_1^2A_2}(\omega_t, \omega_\tau)
\notag
\\
&=
\frac{3}{2}\pi^2 A_1^2A_2 
\Big[
\delta(\omega_t-3\Omega)\delta(\omega_\tau-\Omega)
\chi^{(3)}(3\Omega; \Omega, \Omega, \Omega)
\notag
\\
&\quad
+\delta(\omega_t-\Omega)(2\delta(\omega_\tau-\Omega)+\delta(\omega_\tau+\Omega))
\notag
\\
&\quad\times
\chi^{(3)}(\Omega; \Omega, \Omega, -\Omega)
\notag
\\
&\quad
+\delta(\omega_t+\Omega)(2\delta(\omega_\tau+\Omega)+\delta(\omega_\tau-\Omega))
\notag
\\
&\quad\times
\chi^{(3)}(-\Omega; \Omega, -\Omega, -\Omega)
\notag
\\
&\quad
+\delta(\omega_t+3\Omega)\delta(\omega_\tau+\Omega)
\chi^{(3)}(-3\Omega; -\Omega, -\Omega, -\Omega)
\Big].
\end{align}
In the same way, the signal proportional to $A_1A_2^2$ can be derived from Eq.~(\ref{j A_1A_2^2 formula}) as
\begin{align}
&
j_{A_1A_2^2}(\omega_t, \omega_\tau)
\notag
\\
&=
\frac{3}{2}\pi^2 A_1A_2^2
\notag
\\
&\quad\times
\Big[
\delta(\omega_t-3\Omega)
\delta(\omega_\tau-2\Omega)
\chi^{(3)}(3\Omega; \Omega, \Omega, \Omega)
\notag
\\
&\quad
+\delta(\omega_t-\Omega)
(2\delta(\omega_\tau)+\delta(\omega_\tau-2\Omega))
\chi^{(3)}(\Omega; \Omega, \Omega, -\Omega)
\notag
\\
&\quad
+\delta(\omega_t+\Omega)
(2\delta(\omega_\tau)+\delta(\omega_\tau+2\Omega))
\chi^{(3)}(-\Omega; \Omega, -\Omega, -\Omega)
\notag
\\
&\quad
+\delta(\omega_t+3\Omega)
\delta(\omega_\tau+2\Omega)
\chi^{(3)}(-3\Omega; -\Omega, -\Omega, -\Omega)
\Big].
\end{align}
One can see that the 2DCS signal $j_{\rm NL}(\omega_t, \omega_\tau)
=j_{A_1^2A_2}(\omega_t, \omega_\tau)+j_{A_1A_2^2}(\omega_t, \omega_\tau)$ only appears at several discrete spots.
The components at $(\omega_t/\Omega, \omega_\tau/\Omega)
=(3, 2), (3, 1), (-3, -1), (-3,-2)$ are proportional to
$\chi^{(3)}(3\Omega; \Omega, \Omega, \Omega)$
(or $\chi^{(3)}(-3\Omega; -\Omega, -\Omega, -\Omega)$),
which corresponds to the third harmonic generation.
On the other hand, the components at $(\omega_t/\Omega, \omega_\tau/\Omega)
=(1,2), (1,1)$, $(1,0), (1,-1), (-1,1), (-1,0), (-1,-1), (-1,-2)$
are proportional to $\chi^{(3)}(\Omega; \Omega, \Omega, -\Omega)$
(or $\chi^{(3)}(-\Omega; \Omega, -\Omega, -\Omega)$), which is the susceptibility of the ac Kerr effect.
Hence, the 2DCS in the narrow-band limit is fully characterized by
the THG and ac Kerr susceptibilities, as pointed out previously.
In the literature of 2DCS \cite{Liu2025}, the signals at $(\omega_t, \omega_\tau)=(\Omega, -\Omega), (\Omega, 0), (\Omega, \Omega), (\Omega, 2\Omega)$ are
referred to as rephasing, pump-probe, non-rephasing, and two-quantum responses, respectively.
We summarize the results in the two-dimensional map shown in Fig.~\ref{fig: 2d spectrum narrow-band}.
In this way, 
the 2DCS protocol allows one to isolate nonlinear contributions by combining signals with different pulse configurations. In particular, by subtracting the single-pulse responses $j_1$ and $j_2$ from the double-pulse response $j_{12}$, one can remove linear-response contributions and isolate the cross terms proportional to higher-order susceptibilities. 

\section{Two-dimensional coherent spectroscopy in the broad-band limit}
\label{sec: 2DCS broad-band}

Next, let us consider the opposite limit, i.e., the broad-band limit, in the 2DCS.
As in the case of the linear response, the broad-band pulses are modeled by delta-functions,
\begin{align}
E(t)
&=
A_1 \delta(t)+A_2 \delta(t+\tau),
\end{align}
where the pulse 1 is centered at $t=0$, and the pulse 2 is centered at $t=-\tau$.
The corresponding vector potential is given by $A(t)=A_1(t)+A_2(t+\tau)=-A_1\theta(t)-A_2\theta(t+\tau)$.
The Fourier transform of $A_i(t)$ ($i=1,2$) reads
\begin{align}
A_1(\omega)
&=
-\frac{iA_1}{\omega+i\eta},
\\
A_2(\omega)
&=
-\frac{iA_2}{\omega+i\eta}.
\end{align}

\begin{figure}[t]
\includegraphics[width=8cm]{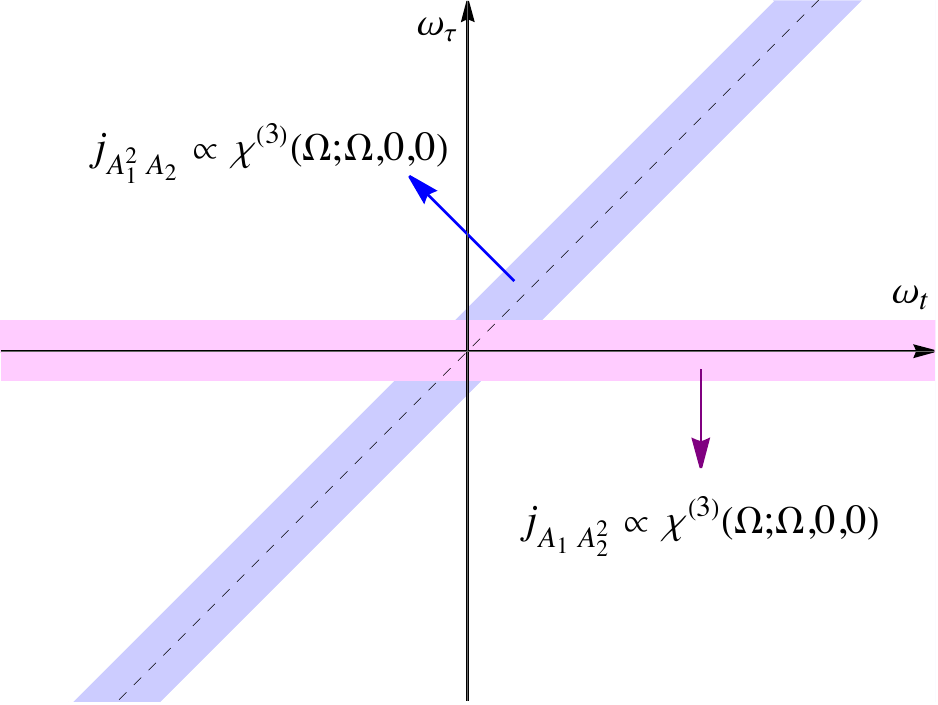}
\caption{2DCS spectrum $j_{\rm NL}(\omega_t, \omega_\tau)$ in the broad-band limit.
The signal is diverging along the diagonal ($\omega_t=\omega_\tau$) and horizontal ($\omega_\tau=0$) lines.
$j_{A_1^2A_2}$ along the diagonal line and $j_{A_1A_2^2}$ along the horizontal line
are proportional to the dc Kerr susceptibility.}
\label{fig: 2d spectrum broad-band}
\end{figure}

From Eq.~(\ref{j A_1^2A_2 formula}), we can evaluate the 2DCS signal proportional to $A_1^2A_2$ in the broad-band limit,
\begin{align}
&
j_{A_1^2A_2}(\omega_t, \omega_\tau)
\notag
\\
&=
\frac{3(-i)^3}{\omega_\tau+i\eta}A_1^2A_2
\int_{-\infty}^\infty \frac{d\omega}{2\pi}
\frac{1}{\omega+i\eta}
\frac{1}{\omega_t-\omega_\tau-\omega+i\eta}
\notag
\\
&\quad\times
\chi^{(3)}(\omega_t; \omega, \omega_t-\omega_\tau-\omega, \omega_\tau).
\end{align}
We can further simplify the expression to get
\begin{align}
&
j_{A_1^2A_2}(\omega_t, \omega_\tau)
\notag
\\
&=
\frac{6(-i)^3}{(\omega_\tau+i\eta)(\omega_t-\omega_\tau+2i\eta)}A_1^2A_2
\notag
\\
&\quad\times
\int_{-\infty}^\infty \frac{d\omega}{2\pi}
\frac{1}{\omega+i\eta}\chi^{(3)}(\omega_t; \omega, \omega_t-\omega_\tau-\omega, \omega_\tau).
\label{j A_1^2A_2 broad-band}
\end{align}
From Eq.~(\ref{j A_1^2A_2 broad-band}), we notice that the signal diverges along the diagonal ($\omega_t=\omega_\tau$)
and horizontal ($\omega_\tau=0$) lines. In Fig.~\ref{fig: 2d spectrum broad-band}, we show the diverging region
in the two-dimensional frequency space. 
In particular, along the diagonal line ($\omega_t=\omega_\tau=\Omega$)
the signal $j_{A_1^2A_2}(\omega_t, \omega_\tau)$ apart from the divergent factor $\frac{1}{\omega_t-\omega_\tau+2i\eta}$ has a simple expression.
To see this, let us write
\begin{align}
j_{A_1^2A_2}(\Omega, \Omega)
&\propto
\frac{6(-i)^3A_1^2A_2}{\Omega}
\int_{-\infty}^\infty \frac{d\omega}{2\pi} 
\frac{1}{\omega+i\eta}\chi^{(3)}(\Omega; \omega, -\omega, \Omega).
\end{align}
Using the relation $\frac{1}{\omega+i\eta}=\mathcal P \frac{1}{\omega}-i\pi\delta(\omega)$
($\mathcal P$ indicates the principal value) and noticing that $\chi^{(3)}(\Omega; \omega, -\omega, \Omega)$
is an even function of $\omega$, we have
\begin{align}
j_{A_1^2A_2}(\Omega, \Omega)
&\propto
\frac{3A_1^2A_2}{\Omega}
\chi^{(3)}(\Omega; \Omega, 0, 0).
\end{align}
Thus, we find that the 2DCS signal proportional to $A_1^2A_2$ along the diagonal line is described by
$\chi^{(3)}(\Omega; \Omega, 0, 0)$, which corresponds to the susceptibility for the dc Kerr effect.
This result is in sharp contrast to the case of the narrow-band limit, where the 2DCS signal is given by
$\chi^{(3)}(\Omega; \Omega, \Omega, \Omega)$ and $\chi^{(3)}(\Omega; \Omega, \Omega, -\Omega)$ (Sec.~\ref{sec: 2DCS narrow-band}).
The lesson from this is that the 2DCS signal is related to different nonlinear susceptibilities depending on the width of the pulses
(narrow-band or broad-band pulses).
We note that the divergent factors in our expressions originate from the idealization of delta-function pulses. 
In realistic situations, pulses have finite duration, and these divergences are regularized by the finite bandwidth of the pulse spectrum.

A similar formula can be derived for the 2DCS signal proportional to $A_1A_2^2$ with the formula (\ref{j A_1A_2^2 formula}).
The result is
\begin{align}
&
j_{A_1A_2^2}(\omega_t, \omega_\tau)
\notag
\\
&=
\frac{6(-i)^3}{(\omega_\tau+i\eta)(\omega_t-\omega_\tau+2i\eta)}A_1A_2^2
\notag
\\
&\quad\times
\int_{-\infty}^\infty \frac{d\omega}{2\pi} 
\frac{1}{\omega+i\eta}
\chi^{(3)}(\omega_t; \omega, \omega_\tau-\omega, \omega_t-\omega_\tau).
\label{j A_1A_2^2 broad-band}
\end{align}
Again, the signal is divergent along the diagonal ($\omega_t=\omega_\tau$) and horizontal ($\omega_\tau=0$) lines.
In the case of $j_{A_1A_2^2}$, further simplification occurs along the horizontal line. In fact,
at $\omega_t=\Omega$ and $\omega_\tau=0$
the susceptibility $\chi^{(3)}(\omega_t; \omega_t-\omega_\tau, \omega, \omega_\tau-\omega)$ becomes an even function of $\omega$,
with which the $\omega$ integral can be performed analytically.
Apart from the diverging factor $\frac{1}{\omega_\tau+i\eta}$, the 2DCS signal is given by
\begin{align}
j_{A_1A_2^2}(\Omega, 0)
&\propto
\frac{3A_1A_2^2}{\Omega}
\chi^{(3)}(\Omega; \Omega, 0, 0).
\end{align}
We find that the 2DCS signal proportional to $A_1A_2^2$ along the horizontal line is described by
the dc Kerr susceptibility $\chi^{(3)}(\Omega; \Omega, 0, 0)$.

If we add the contributions proportional to $A_1^2A_2$ and $A_1A_2^2$, we obtain the expression for the total nonlinear current
along the diagonal line,
\begin{align}
j_{\rm NL}(\Omega, \Omega)
&\propto
\frac{3}{\Omega}(A_1^2A_2+A_1A_2^2)
\chi^{(3)}(\Omega; \Omega, 0, 0)
\notag
\\
&\quad
+\frac{6(-i)^3}{\Omega}A_1A_2^2
\notag
\\
&\quad\times
\mathcal P\int_{-\infty}^\infty \frac{d\omega}{2\pi} 
\frac{1}{\omega}
\chi^{(3)}(\Omega; \omega, \Omega-\omega, 0),
\label{j_NL diagonal}
\end{align}
and along the horizontal line,
\begin{align}
j_{\rm NL}(\Omega, 0)
&\propto
\frac{3}{\Omega}(A_1^2A_2+A_1A_2^2)
\chi^{(3)}(\Omega; \Omega, 0, 0)
\notag
\\
&\quad
+\frac{6(-i)^3}{\Omega}A_1^2A_2
\notag
\\
&\quad\times
\mathcal P\int_{-\infty}^\infty \frac{d\omega}{2\pi}
\frac{1}{\omega}\chi^{(3)}(\Omega; \omega, \Omega-\omega, 0).
\label{j_NL horizontal}
\end{align}
Here we have the frequency integral of the third-order nonlinear susceptibility on top of the dc Kerr susceptibility.

To summarize, the 2DCS in the broad-band limit has divergent signals along the diagonal and horizontal lines.
If one keeps track of the intensity dependence ($A_1^2A_2$ or $A_1A_2^2$), 
the signal is proportional to the dc Kerr susceptibility $\chi^{(3)}(\Omega; \Omega, 0, 0)$.
For other regions in the two-dimensional frequency space, the signal is described by
the complicated integral of the third-order nonlinear susceptibility (see the formulas (\ref{j A_1^2A_2 broad-band})
and (\ref{j A_1A_2^2 broad-band})). In the next section, we numerically evaluate the ac and dc Kerr susceptibilities
for a model of disordered superconductors.

\begin{figure*}[t]
\includegraphics[width=16cm]{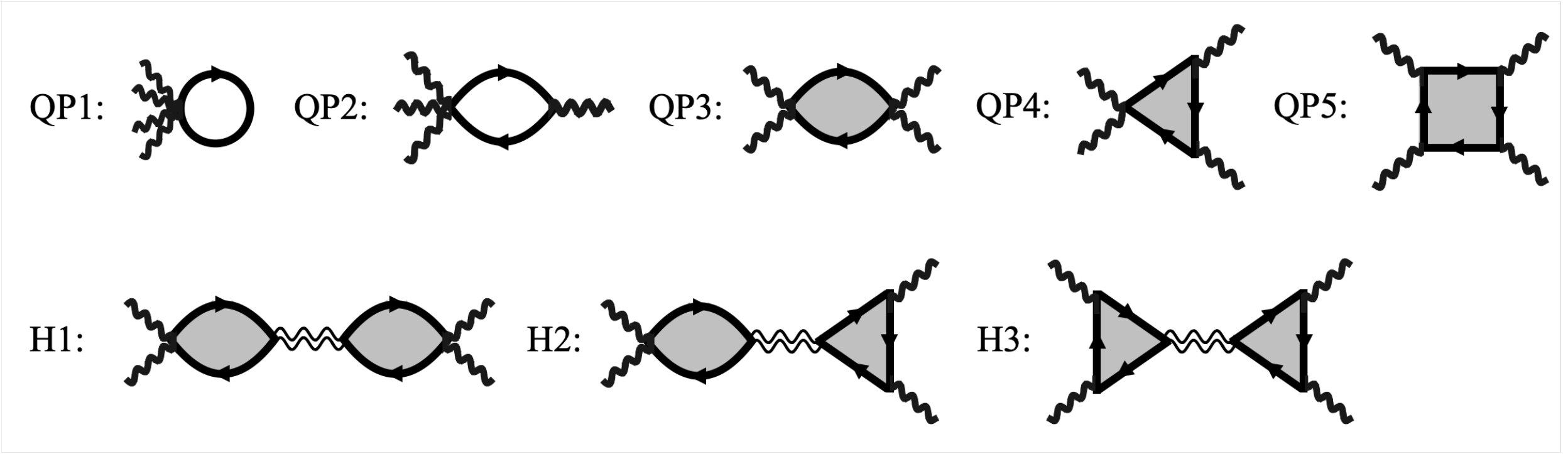}
\caption{
Classification of the third-order susceptibility diagrams for disordered superconductors within the self-consistent Born approximation 
(see also Table II in Ref.~\cite{TsujiNomura2020}).
The single wavy lines, double wavy lines, and solid lines represent the photon, Higgs-mode, and electron propagators, respectively. 
The shaded triangles and boxes indicate electron loop diagrams including impurity corrections.
The labels QP1, $\dots$, QP5 denote the quasiparticle contributions, while H1, $\dots$, H3 denote the Higgs-mode contributions.}
\label{fig: kerr diagram full}
\end{figure*}

\section{Numerical simulation}
\label{sec: numerical}

\subsection{Model}

In the previous sections, we have seen that the 2DCS signals in the narrow-band and broad-band limits are related to
the ac and dc Kerr susceptibilities ($\chi^{(3)}(\Omega; \Omega, \Omega, -\Omega)$ and $\chi^{(3)}(\Omega; \Omega, 0, 0)$).
In this section, we evaluate them numerically for a lattice model of disordered superconductors. The Hamiltonian is defined by
\begin{align}
\hat H(t)
&=
\sum_{\bm k, \sigma} \epsilon_{\bm k-\bm A(t)} c_{\bm k\sigma}^\dagger c_{\bm k\sigma}
-\frac{V}{N} \sum_{\bm k, \bm k'} c_{\bm k\uparrow}^\dagger c_{-\bm k\downarrow}^\dagger
c_{-\bm k'\downarrow} c_{\bm k'\uparrow}
\notag
\\
&\quad
+\sum_{i, \sigma} v_i c_{i\sigma}^\dagger c_{i\sigma},
\label{Hamiltonian}
\end{align}
where $c_{\bm k\sigma}^\dagger$ is the creation operator of electrons with momentum $\bm k$ and spin $\sigma$,
$\epsilon_{\bm k}$ is the band dispersion, $V(>0)$ is the attractive interaction strength, $N$ is the number of $k$ points,
and $v_i$ is a local random potential at site $i$. 

We assume the BCS pairing interaction in the $s$-wave channel,
for which we employ the mean-field approximation to decompose the interaction term. The gap function is defined by
\begin{align}
\Delta(t)
&=
-\frac{V}{N}\sum_{\bm k} \langle c_{-\bm k\downarrow}(t) c_{\bm k\uparrow}(t)\rangle.
\label{gap eq}
\end{align}
Without loss of generality, we take $\Delta(t)$ to be a real value \cite{phase_mode}.
The disorder potential $v_i$  is assumed to be
a Gaussian random variable with the disorder average, $\langle v_i v_j\rangle_{\rm disorder}=\gamma^2 \delta_{ij}$.
Here $\gamma$ is the average impurity scattering rate. Within the self-consistent Born approximation,
the effect of the disorder is included in the form of the self-energy \cite{AbrikosovGorkov1959, TsujiNomura2020},
\begin{align}
\Sigma(t,t')
&=
\frac{\gamma^2}{N}\sum_{\bm k} G_{\bm k}(t,t'),
\end{align}
where $G_{\bm k}(t,t')$ is the single-particle Green's function in the Keldysh formalism \cite{Keldysh1964, Aoki2014}.
In Appendix \ref{app: Green's function}, we briefly summarize the Keldysh formalism
for the Green's function.

In our numerical simulation, we consider the square lattice with the dispersion $\epsilon_{\bm k}=-2t_h(\cos k_x+\cos k_y)$
($t_h$ is the hopping amplitude and we set the lattice constant $a=1$), 
which has the inversion symmetry. We use $t_h$ as the unit of energy, and set $t_h=1$ throughout the paper.
The pulse field is assumed to be polarized along the $x$ axis (i.e., $\bm e=(1,0)$). The number of particles is fixed at half filling (the chemical potential is set to $\mu=0$).
The number of $k$ points used in this paper is taken to be $N=100\times 100$.
We confirm that the results do not change even if we increase $N$ up to $200\times 200$.
We use the interaction parameter $V=2.5$, and take a small imaginary part $\eta=0.01$ as a broadening factor
for the Green's function and nonlinear susceptibilities.

We solve the Green's function self-consistently within the BCS and self-consistent Born approximations.
The nonlinear susceptibilities are systematically derived by differentiating the current,
\begin{align}
j(t)
&=
\sum_{\bm k\sigma\mu} e_\mu \frac{\partial\epsilon_{\bm k-\bm A(t)}}{\partial k_\mu} \langle c_{\bm k\sigma}^\dagger (t)c_{\bm k\sigma}(t)\rangle,
\end{align}
with respect to the driving field $A(t)$. The derivation is analogous to that of the third harmonic generation \cite{Tsuji2016, TsujiNomura2020}.
The nonlinear susceptibility is subject to the impurity and Higgs-mode vertex corrections, which should be included consistently with the form of the self-energy.
The details of the impurity and Higgs-mode vertices are summarized in Appendix \ref{app: vertex correction}.
The explicit form of the nonlinear susceptibilities is given in Appendix \ref{app: chi3}.

\begin{table}
\caption{
The number of photon lines attached to each vertex and the type of the coupling to electromagnetic fields
in the third-order nonlinear susceptibility diagrams (Fig.~\ref{fig: kerr diagram full}) for disordered superconductors within the self-consistent Born approximation.
For the Higgs-mode contributions, the vertices are split into two groups separated by the Higgs-mode propagator.
}
\label{table: classification}
\begin{tabular}{ccc}
\hline
\hline
\; label \; & number of photons & coupling
\\
\hline
QP1 & $4$ & diamag
\\
QP2 & $3+1$ & paramag
\\
QP3 & $2+2$ & diamag
\\
QP4 & $2+1+1$ & diamag + paramag
\\
QP5 & $1+1+1+1$ & paramag
\\
H1 & $(2)+(2)$ & diamag
\\
H2 & $(2)+(1+1)$ & diamag + paramag
\\
H3 & $(1+1)+(1+1)$ & paramag
\\
\hline
\hline
\end{tabular}
\end{table}

The nonlinear susceptibility diagrams for disordered superconductors are decomposed into the quasiparticle and Higgs-mode contributions,
each of which is further classified according to the number of photon lines attached to each vertex.
We show the classified diagrams in Fig.~\ref{fig: kerr diagram full}.
There are five types of diagrams assigned to the contribution of quasiparticles (QP1, ..., QP5)
and three types of diagrams assigned to the Higgs-mode contribution (H1, ..., H3).
For the latter, the photon vertices are split into two groups
separated by the Higgs-mode propagator.
Within the self-consistent Born approximation, the inversion symmetry is effectively recovered if the original system without disorders has the inversion symmetry. 
In this case, the number of photon lines in each group must be even since
vertices with the odd number of photons are parity odd and the momentum integral of an odd parity function vanishes.
For example, grouping such as $(3)+(1)$, $(2+1)+(1)$, and $(1+1+1)+(1)$ does not appear
for the Higgs-mode diagram.
The coupling to electric fields is either diamagnetic or paramagnetic
depending on an even or odd number of photons attached to the vertex.
We summarize the number of photon lines and the coupling type for each susceptibility diagram
in Table \ref{table: classification}.
In the study of third harmonic generation \cite{TsujiNomura2020}, it has been known that the diagrams with the paramagnetic coupling
become dominant over those with the diamagnetic coupling if the system is in the dirty regime.
We will see that similar behavior is observed for the 2DCS signal.

\subsection{Results}

\begin{figure}[t]
\includegraphics[width=8cm]{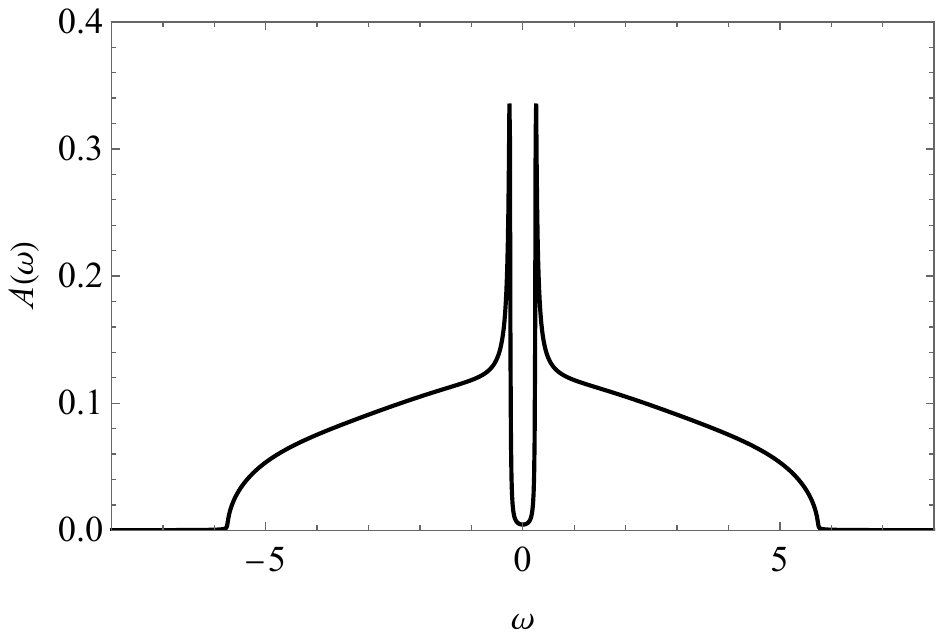}
\caption{Single-particle spectrum for the lattice model of disordered superconductors in the dirty regime.
The parameters are $V=2.5, \gamma=2$, and $\beta=50$.
The gap size is $2\Delta=0.50$ and $\gamma/(2\Delta)=4.0$.}
\label{fig: A(w)}
\end{figure}

\begin{figure}[t]
\includegraphics[width=8cm]{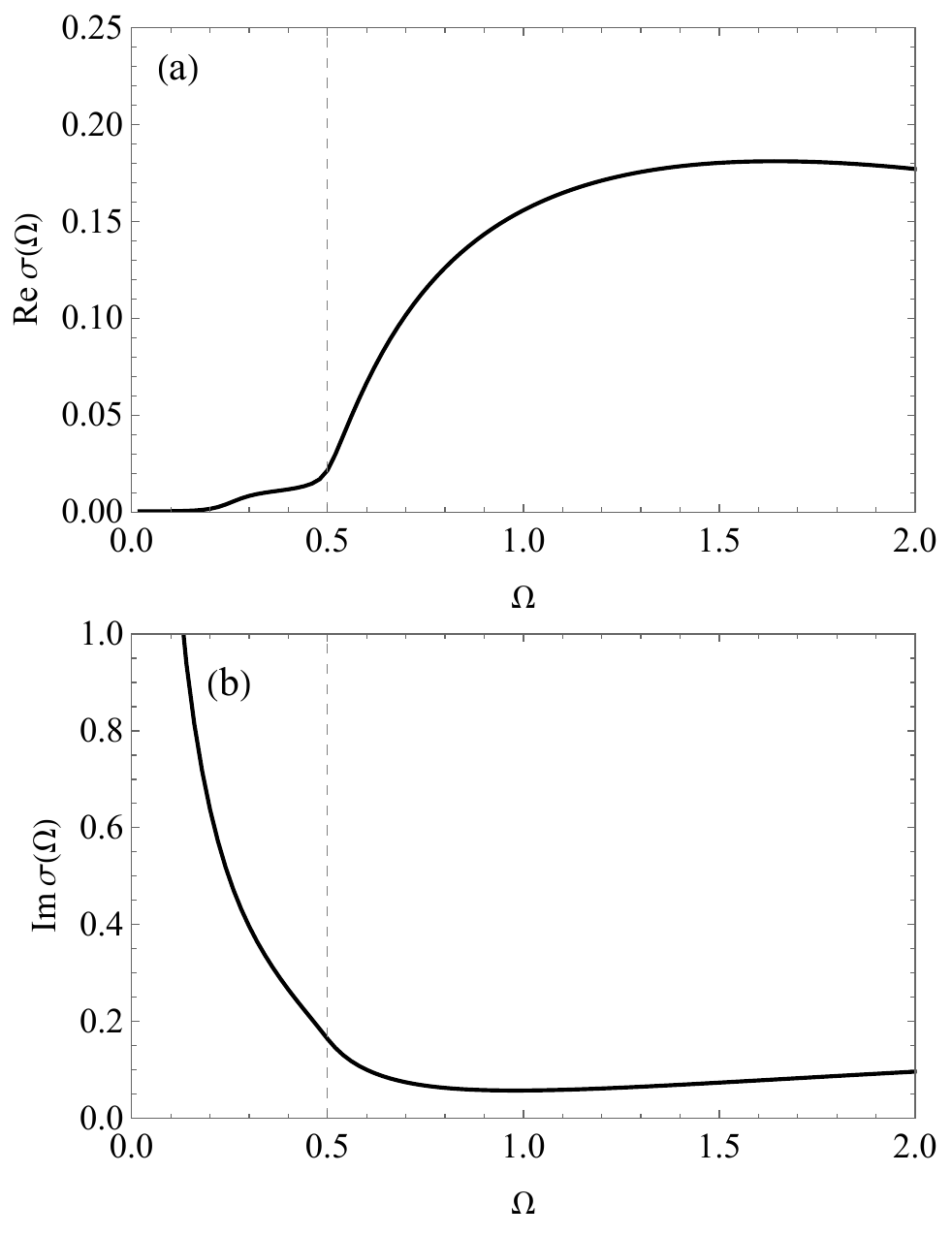}
\caption{(a) Real and (b) imaginary parts of the optical conductivity $\sigma(\Omega)$ for the lattice model of disordered superconductors in the dirty regime. The parameters are $V=2.5, \gamma=2$, and $\beta=50$.
The dashed lines indicate the gap frequency $2\Delta=0.50$.}
\label{fig: optical conductivity}
\end{figure}

Let us first look at the equilibrium and linear-response spectra for our model.
Figure \ref{fig: A(w)} shows the single-particle spectrum $A(\omega)$ in the superconducting phase of the lattice model
with $V=2.5$, $\gamma=2$, and the inverse temperature $\beta=50$. As we will see below, the superconducting transition temperature is $T_c\approx 0.14$
($\beta_c\approx 7.1$) in this case, so that the system is in the superconducting phase.
In the single-particle spectrum, we can see that a gap is opening with $2\Delta=0.50$ at $\beta=50$. 
This means that the system is in the dirty regime
($\gamma/(2\Delta)=4.0$). We can confirm that the relation (\ref{scale separation}) is roughly satisfied ($W=8, \gamma=2, 2\Delta=0.50, v_F/L=0.02$).

In Fig.~\ref{fig: optical conductivity}, we plot the optical conductivity $\sigma(\Omega)$ for the same parameter set.
The real part [Fig.~\ref{fig: optical conductivity}(a)] shows a build-up of the spectral weight above the gap frequency, resembling 
the Mattis-Bardeen form of the optical conductivity in the dirty limit \cite{MattisBardeen1958}. 
The result is consistent with the fact that the system is in the dirty regime ($\gamma/(2\Delta)=4.0$).
The spectral weight below the gap apprears due to the presence of the small broadening factor ($\eta=0.01$) in the Green's function.
The subgap weight vanishes in the limit of $\eta\to 0$ (and $T\to 0$) \cite{Zimmermann1991}.
The imaginary part [Fig.~\ref{fig: optical conductivity}] shows
a $1/\Omega$ divergence at low frequency, which signals a nonzero superfluid stiffness.

\begin{figure}[t]
\includegraphics[width=8cm]{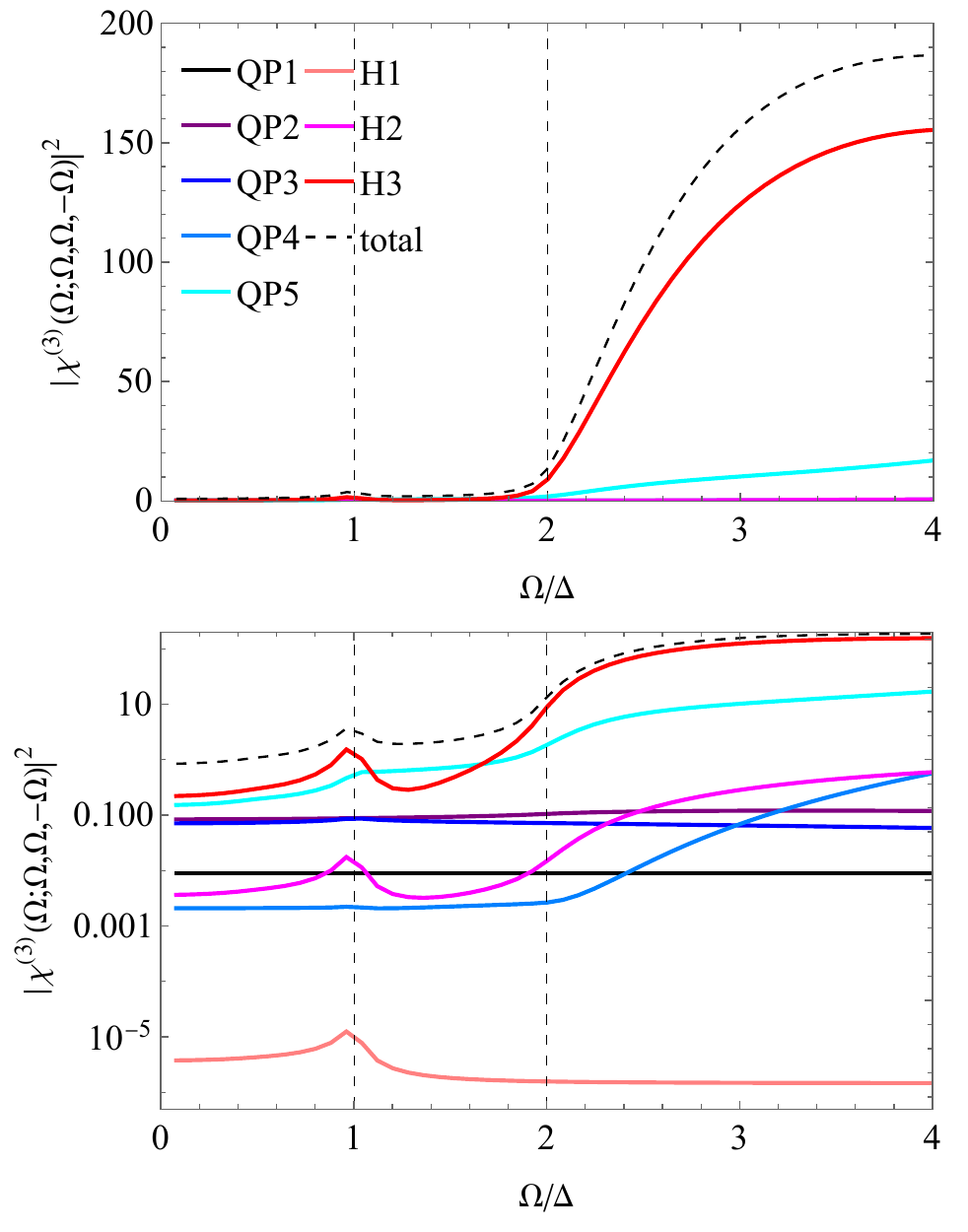}
\caption{Intensity of the ac Kerr susceptibility $\chi^{(3)}(\Omega; \Omega, \Omega, -\Omega)$ as a function of $\Omega/\Delta$
in the linear scale (top panel) and log scale (bottom)
for the lattice model of disordered superconductors in the dirty regime. The parameters are $V=2.5$, $\gamma=2$, and $\beta=50$.
The vertical dashed lines indicate $\Omega=\Delta$ and $\Omega=2\Delta$.}
\label{fig: ac kerr}
\end{figure}

We then move on to see the ac Kerr susceptibility $\chi^{(3)}(\Omega; \Omega, \Omega, -\Omega)$, which describes the 2DCS signal
in the narrow-band limit (Sec.~\ref{sec: 2DCS narrow-band}).
In Fig.~\ref{fig: ac kerr}, we plot $|\chi^{(3)}(\Omega; \Omega, \Omega, -\Omega)|^2$ as a function of $\Omega/\Delta$
in the linear scale (top panel) and log scale (bottom)
in the superconducting phase of the lattice model ($V=2.5$, $\gamma=2$, and $\beta=50$).
The total intensity is shown by the dashed curve in Fig.~\ref{fig: ac kerr}.
We classify the total contribution into eight different diagrams (QP1, $\dots$, QP5, H1, $\dots$, H3) as in Table \ref{table: classification}.
We can see that $\chi^{(3)}(\Omega; \Omega, \Omega, -\Omega)$ shows threshold behavior in the linear scale, i.e., the spectral weight 
starts to grow at the gap frequency $\Omega=2\Delta$ as $\Omega$ increases.
This is in contrast to the previous argument \cite{Katsumi2024}
that $\chi^{(3)}(\Omega; \Omega, \Omega, -\Omega)$ could be resonantly enhanced at $\Omega=2\Delta$.
We attribute this to the difference of the treatments of disorders applied to different parameter regimes (see also the discussion in Sec.~\ref{sec: introduction}).
In Ref.~\cite{Katsumi2024}, the calculation is performed on relatively small lattices, 
where it may be difficult to realize the scale separation represented by the inequality (\ref{scale separation}).
In the Born regime that we are focusing on, the impurity scattering rate should be sufficiently small compared to the hopping amplitude but should be sufficiently large compared to the superconducting gap size.

At $\Omega\ge 2\Delta$, the dominant contribution to the ac Kerr susceptibility is coming from the Higgs mode with the paramagnetic coupling (H3 diagram).
The second dominant contribution is from the QP5 diagram (i.e., quasiparticles with the paramagnetic coupling to single photons).
The other diagrams have much smaller weights with the difference of several orders of magnitude.
This is similar to the behavior of third harmonic generation that shows resonance at $\Omega=\Delta$ \cite{Tsuji2015, TsujiNomura2020}.
The THG susceptibility is also dominated by the Higgs mode in the dirty regime.
In Appendix \ref{app: thg}, we show the results of the THG susceptibility for the same model as a comparison.

\begin{figure}[t]
\includegraphics[width=6cm]{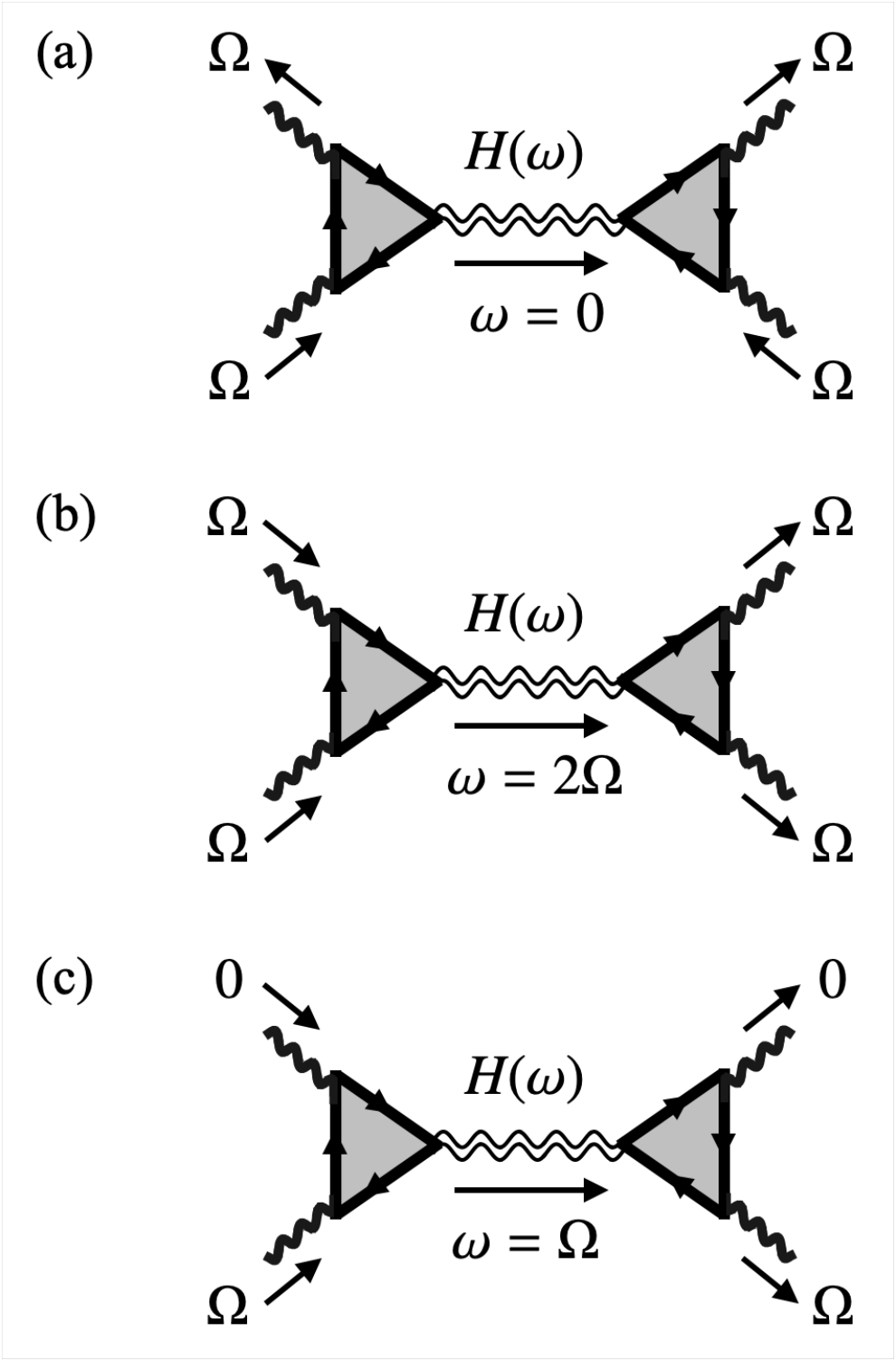}
\caption{H3 diagrams for [(a), (b)] the ac Kerr susceptibility $\chi^{(3)}(\Omega; \Omega, \Omega, -\Omega)$
with the Higgs-mode propagator $H(\omega)$ (double wavy lines) carrying frequency (a) $\omega=0$ and (b) $\omega=2\Omega$,
and (c) the dc Kerr susceptibility $\chi^{(3)}(\Omega; \Omega, 0, 0)$. The single wavy lines and solid lines represent
the photon propagators and electron propagators, respectively, while the shaded triangles indicate electron loop diagrams including impurity corrections.}
\label{fig: kerr diagram}
\end{figure}

\begin{figure}[t]
\begin{center}
\includegraphics[width=8cm]{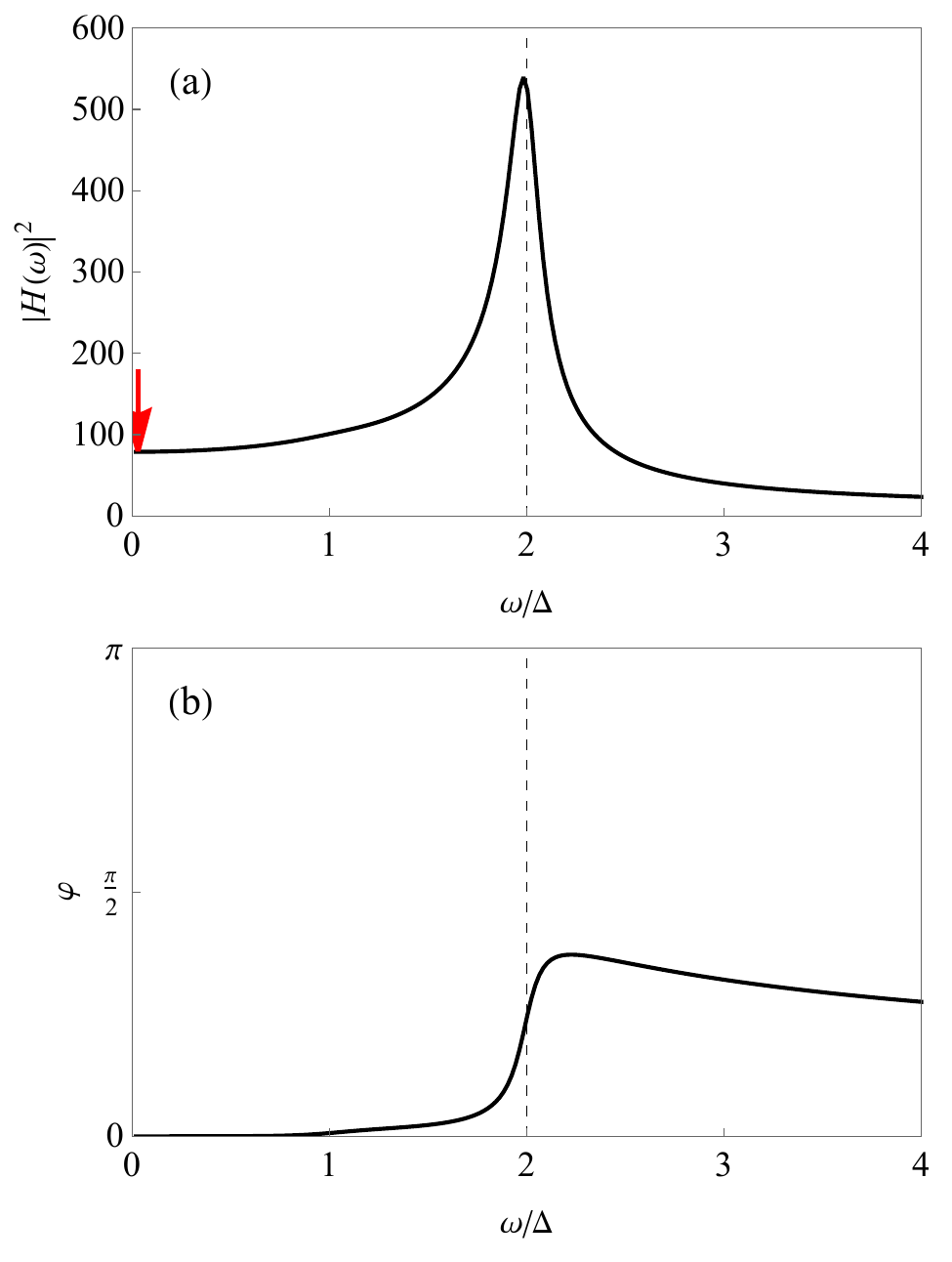}
\caption{
(a) The intensity and (b) the phase of the Higgs-mode propagator $H(\omega)$ as a function of $\omega/\Delta$
for the lattice model of disordered superconductors in the dirty regime.
The parameters are $V=2.5$, $\gamma=2$, and $\beta=50$.
The red arrow shows a spectral weight at $\omega=0$.}
\label{fig: higgs propagator}
\end{center}
\end{figure}

The interpretation of these results is as follows: The H3 diagrams for $\chi^{(3)}(\Omega; \Omega, \Omega, -\Omega)$
are shown in Fig.~\ref{fig: kerr diagram}(a) and (b) depending on the combination of the in-coming and out-going photons. 
Each nonlinear process is mediated by the Higgs-mode propagator $H(\omega)$ as shown by the wavy lines in Fig.~\ref{fig: kerr diagram}.
For $\Omega\ge 2\Delta$, the main contribution comes from the diagram (a), since (b) is suppressed due to
high frequency ($2\Omega\ge 4\Delta$) carried by the Higgs-mode propagator $H(\omega)$ (for the definition, see Appendix \ref{app: vertex correction}. 
We plot the intensity and phase of $H(\omega)$ in Fig.~\ref{fig: higgs propagator}, which shows a sharp peak structure
and an approximate $\pi/2$ phase shift at $\omega=2\Delta$.
In Fig.~\ref{fig: kerr diagram}(a), the Higgs mode carries zero frequency, meaning that 
the process is far off-resonant (recall that the eigenfrequency of the Higgs mode at zero momentum is $2\Delta$). 
In fact, the Higgs-mode propagator has a nonzero weight at $\omega=0$ (the red arrow in Fig.~\ref{fig: higgs propagator}(a)).
Thus, the ac Kerr susceptibility in charge of the 2DCS in the narrow-band limit is mainly mediated by the zero-frequency ($\omega=0$) Higgs mode,
picking up the tail of the Higgs-mode resonance peak centered at $\omega=2\Delta$.
This scenario is similar to the one employed for the pump-probe spectroscopy of superconductors in the clean regime \cite{ShimanoTsuji2020, Katsumi2018}.
In the ac Kerr diagram with the zero-frequency Higgs mode (Fig.~\ref{fig: kerr diagram}(a)), the Higgs-mode propagator does not depend on $\Omega$ but becomes a constant. 
Hence the diagram is effectively split into two triangle parts, each of which resembles the optical conductivity diagram.
As a result, $\chi^{(3)}(\Omega; \Omega, \Omega, -\Omega)$ shows a similar frequency dependence (i.e., threshold behavior) at $\Omega=2\Delta$.

If we look at the log-scale plot of $|\chi^{(3)}(\Omega; \Omega, \Omega, -\Omega)|^2$ (bottom panel of Fig.~\ref{fig: ac kerr}), 
we notice that there is a tiny resonance peak at $\Omega=\Delta$ (half of the gap frequency),
which is the same resonance frequency as in THG. This resonance is also dominated by the H3 diagram,
and can be simply explained by the two-photon absorption process (see also Ref.~\cite{Jujo2015}). 
As shown in Fig.~\ref{fig: kerr diagram}(b), the Higgs mode can carry the frequency $\omega=2\Omega$ due to two-photon absorption,
and resonance occurs at $\Omega=\Delta$, where $2\Omega$ agrees with the eigenfrequency of the Higgs mode.
This is similar to the story of the THG resonance observed at $\Omega=\Delta$, in which the H3 diagram also plays the dominant role (see also Appendix \ref{app: thg}).

\begin{figure}[t]
\includegraphics[width=8cm]{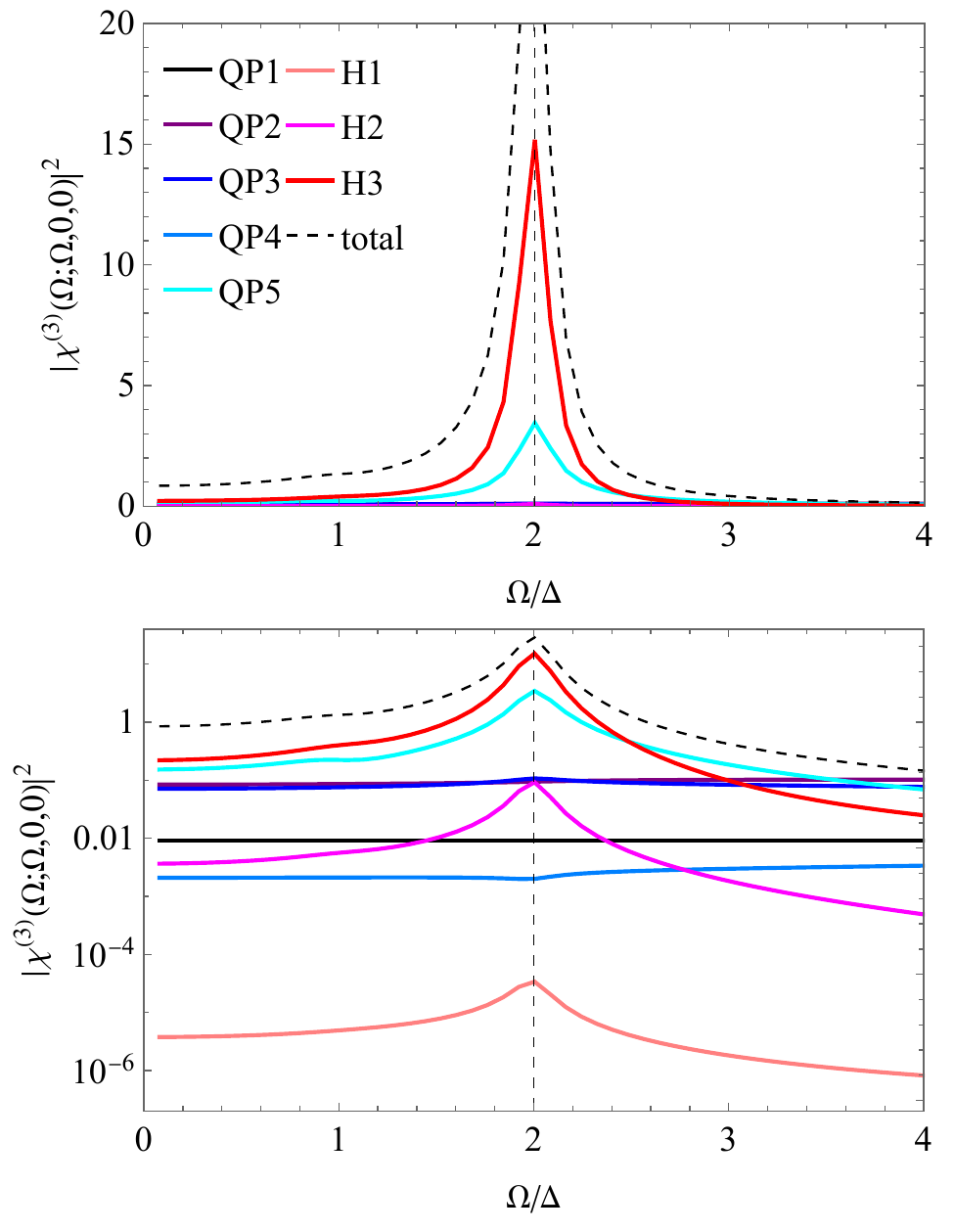}
\caption{Intensity of the dc Kerr susceptibility $\chi^{(3)}(\Omega; \Omega, 0, 0)$ as a function of $\Omega/\Delta$
in the linear scale (top panel) and log scale (bottom)
for the lattice model of disordered superconductors in the dirty regime. The parameters are $V=2.5$, $\gamma=2$, and $\beta=50$.
The vertical dashed lines indicate $\Omega=2\Delta$.}
\label{fig: dc kerr}
\end{figure}

Let us comment on the lifetime and decaying behavior of the Higgs mode.
The amplitude of the Higgs-mode propagator shows a clear peak structure with a finite peak height (Fig.~\ref{fig: higgs propagator}(a)). 
The width of the peak roughly gives the decay rate of the Higgs mode due to impurity scattering 
(in the clean limit the propagator diverges as $(\omega-2\Delta)^{-1/2}$ and the Higgs mode decays in a power law).
However, the phase of the propagator approximately shows a $\pi/2$ shift (Fig.~\ref{fig: higgs propagator}b) as in the clean limit, 
suggesting that the decay is possibly a combination of a power-law decay to quasiparticles in short time and an exponential decay due to impurity scattering
in a longer time scale.

Next, let us examine the numerical results for the dc Kerr susceptibility $\chi^{(3)}(\Omega; \Omega, 0, 0)$,
which is related to the 2DCS signal along the diagonal and horizontal lines in the broad-band limit (Sec.~\ref{sec: 2DCS broad-band}).
In Fig.~\ref{fig: dc kerr}, we plot $|\chi^{(3)}(\Omega; \Omega, 0, 0)|^2$ as a function of $\Omega/\Delta$ in the linear scale (top panel)
and log scale (bottom) in the superconducting phase of the lattice model ($V=2.5$, $\gamma=2$, and $\beta=50$).
The total intensity is shown by the dashed curve in Fig.~\ref{fig: dc kerr}.
We decompose the total contribution into different diagrams (QP1, $\dots$, QP5, H1, $\dots$, H3 as in Table \ref{table: classification}).
We can observe that the dc Kerr susceptibility shows a resonance peak at $\Omega=2\Delta$, which is dominated by the H3 diagram.
The second largest contribution comes from the QP5 diagram.
The fact that the H3 diagram plays a dominant role is similar to the case of the ac Kerr susceptibility (Fig.~\ref{fig: ac kerr}) and THG susceptibility \cite{TsujiNomura2020} (Appendix \ref{app: thg})
in the dirty regime of superconducting systems. 
The reason that the resonance appears at $\Omega=2\Delta$ can be understood from
the H3 diagram shown in Fig.~\ref{fig: kerr diagram}(c). 
In the case of the dc Kerr effect, there are an in-coming photon with frequency $\Omega$ and another in-coming photon with zero frequency.
In total, the Higgs-mode propagator carries the fundamental driving frequency $\Omega$, with which the resonance occurs when $\Omega$ agrees with
the eigenfrequency of the Higgs mode $2\Delta$. 
Thus, the frequency dependence of the 2DCS signal may become qualitatively different between the narrow-band 
and broad-band limit: In the former, the signal grows towards high frequencies with the threshold $\Omega=2\Delta$. In the latter,
the signal along the diagonal and horizontal lines is partly described by the dc Kerr susceptibility, which shows
the resonance peak at $\Omega=2\Delta$. In realistic situations, the behavior of the 2DCS signal will be
in between the narrow-band and broad-band limits, which may help to understand the experimental results.

\begin{figure}[t]
\includegraphics[width=8cm]{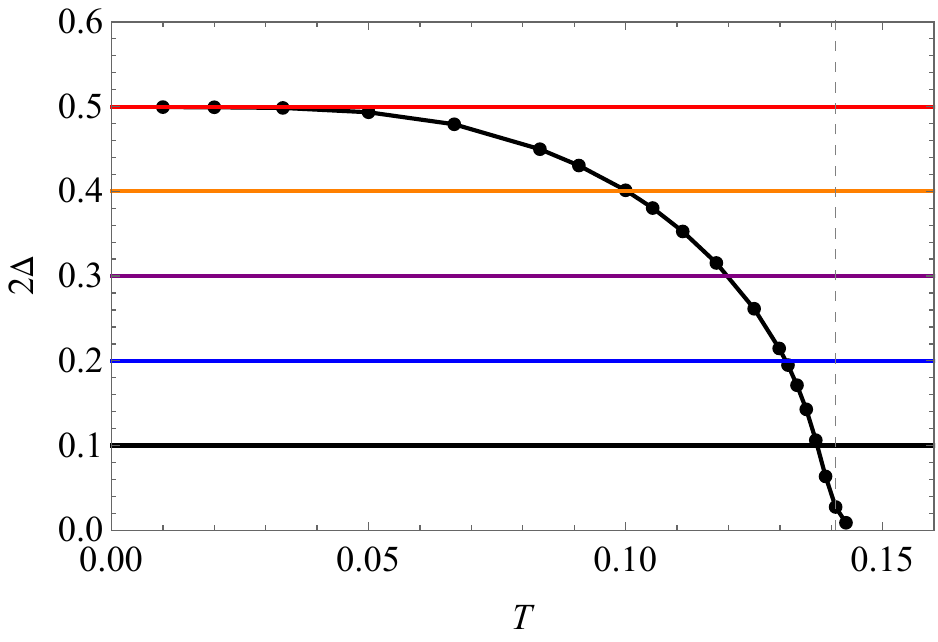}
\caption{Temperature dependence of the superconducting gap $2\Delta$ for the lattice model of disordered superconductors with $V=2.5$ and $\gamma=2$.
The dashed line shows the critical temperature $T_c\approx 0.14$. The horizontal lines represent constant frequencies $\Omega=0.1, 0.2, \dots, 0.5.$}
\label{fig: gap T}
\end{figure}

To get closer to experiments, we study the temperature dependence of the ac and dc Kerr susceptibilities of disordered superconductors.
In Fig.~\ref{fig: gap T}, we show the temperature dependence of the equilibrium superconducting gap $2\Delta$ for $V=2.5$ and $\gamma=2$.
The gap grows below the critical temperature $T_c\approx 0.14$. We overlay constant frequency lines $\Omega=0.1, 0.2, \dots, 0.5$ in Fig.~\ref{fig: gap T}.
The crossing between the $2\Delta$ curve and horizontal lines corresponds to the point where $\Omega=2\Delta$ is fulfilled.

\begin{figure}[t]
\includegraphics[width=8cm]{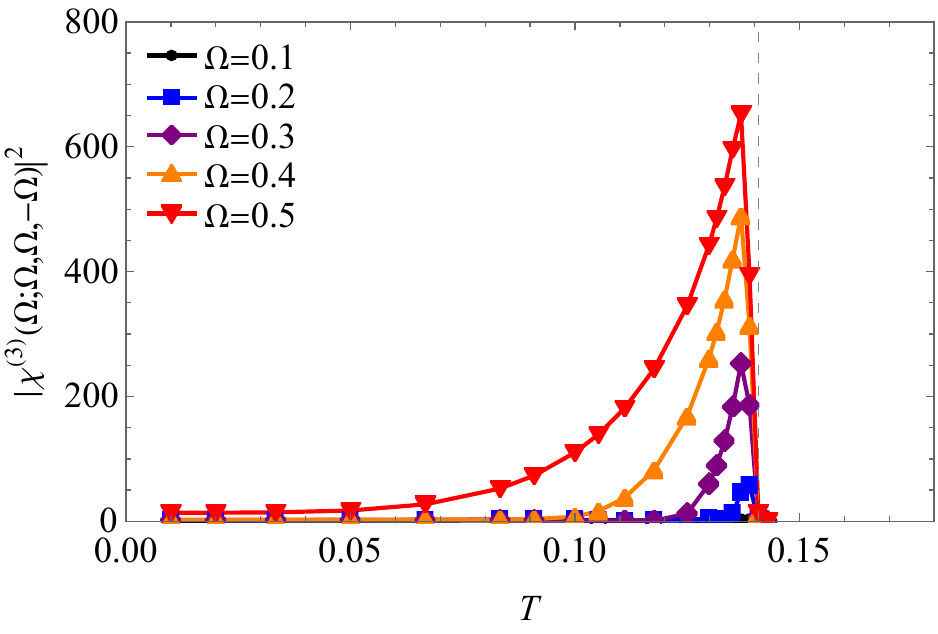}
\caption{Temperature dependence of the ac Kerr susceptibility $\chi^{(3)}(\Omega; \Omega, \Omega, -\Omega)$
at $\Omega=0.1, 0.2, \dots, 0.5$ 
for the lattice model of disordered superconductors in the dirty regime with $V=2.5$ and $\gamma=2$.
The dashed line shows the critical temperature $T_c\approx 0.14$.}
\label{fig: ac Kerr T}
\end{figure}

\begin{figure}[t]
\includegraphics[width=8cm]{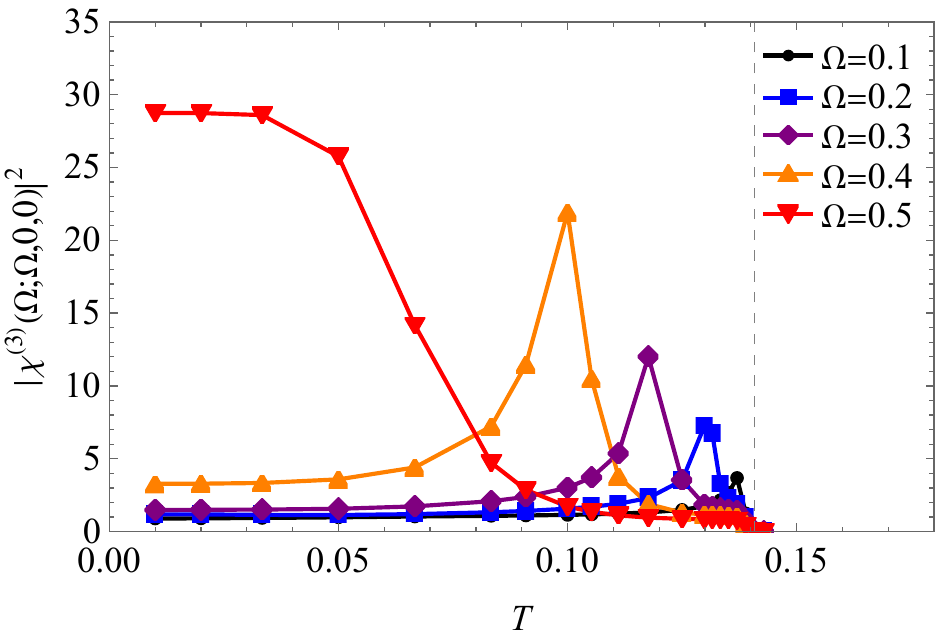}
\caption{Temperature dependence of the dc Kerr susceptibility $\chi^{(3)}(\Omega; \Omega, 0, 0)$
at $\Omega=0.1, 0.2, \dots, 0.5$ 
for the lattice model of disordered superconductors in the dirty regime with $V=2.5$ and $\gamma=2$.
The dashed line shows the critical temperature $T_c\approx 0.14$.}
\label{fig: dc Kerr T}
\end{figure}

In Fig.~\ref{fig: ac Kerr T}, we plot the temperature dependence of the intensity of the ac Kerr susceptibility $|\chi^{(3)}(\Omega; \Omega, \Omega, -\Omega)|^2$
at $\Omega=0.1, 0.2, \dots, 0.5$. As temperature decreases, the susceptibility sharply develops at $T=T_c$, and then gradually decreases toward
low temperature. 
There seems to be a peak in the temperature profile, but this is not caused by resonance with the Higgs mode.
In fact, the peak position does not change as $\Omega$ is varied but stays in the vicinity of $T_c$ irrespective of $\Omega$.
This behavior can be understood from the previous observation that the ac Kerr susceptibility has the threshold at $\Omega=2\Delta$.
When temperature decreases below $T_c$, the spectral weight of $\chi^{(3)}(\Omega; \Omega, \Omega, -\Omega)$ rapidly builds up for a wide range of $\Omega$.
As temperature goes down further, there appears a spectral gap at $\Omega\le 2\Delta$. If we fix $\Omega$ and decrease $T$, 
the ac Kerr susceptibility gets suppressed when $2\Delta(T)\ge \Omega$ (see Fig.~\ref{fig: gap T}), 
explaining the behavior of $\chi^{(3)}(\Omega; \Omega, \Omega, -\Omega)$ at low temperature.

\begin{figure}[t]
\includegraphics[width=8cm]{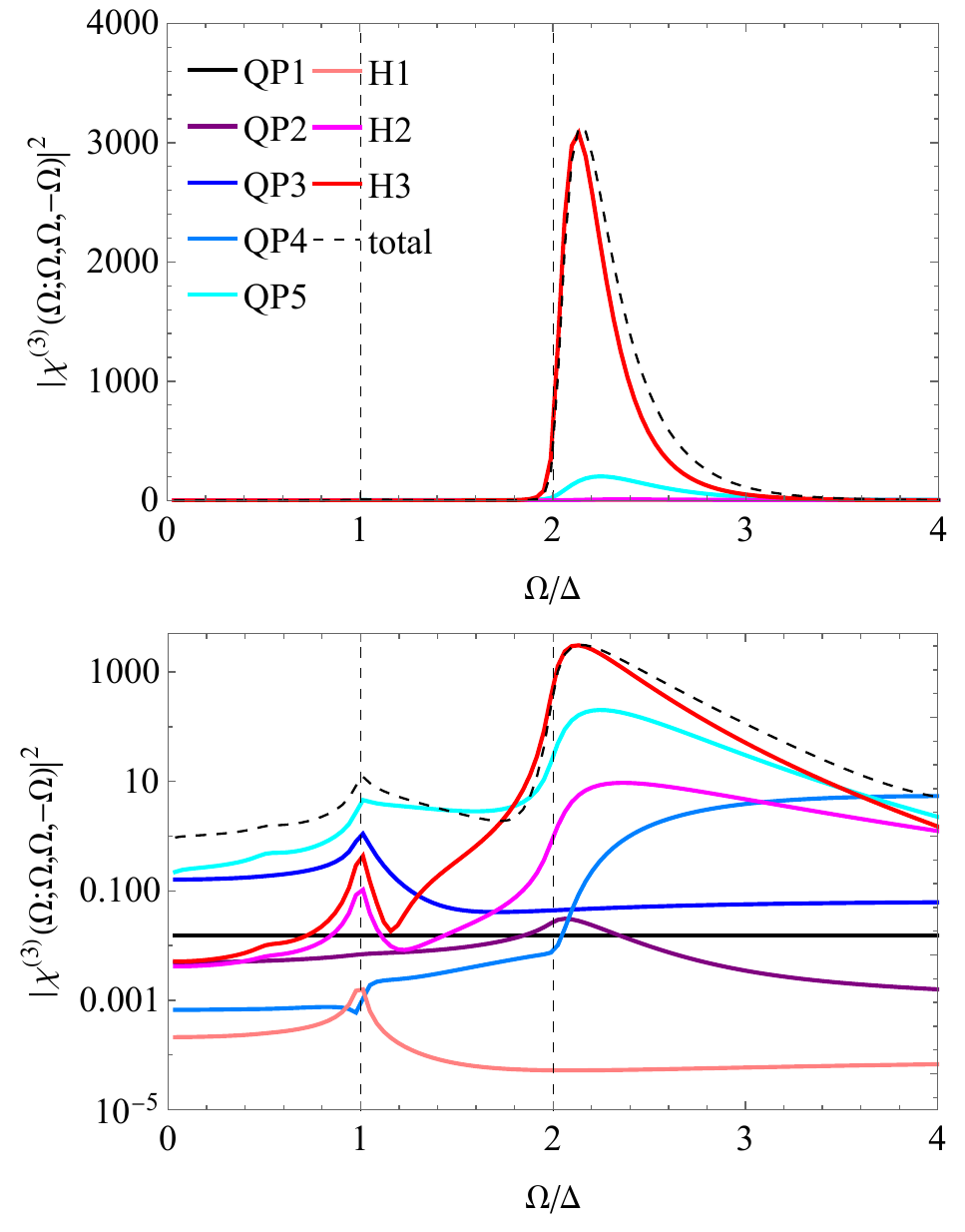}
\caption{Intensity of the ac Kerr susceptibility $\chi^{(3)}(\Omega; \Omega, \Omega, -\Omega)$ as a function of $\Omega/\Delta$
in the linear scale (top panel) and log scale (bottom)
for the lattice model of disordered superconductors in the clean regime. The parameters are $V=2.5$, $\gamma=0.5$, and $\beta=50$.
The vertical dashed lines indicate $\Omega=\Delta$ and $\Omega=2\Delta$.}
\label{fig: ac kerr g05}
\end{figure}

\begin{figure}[t]
\includegraphics[width=8cm]{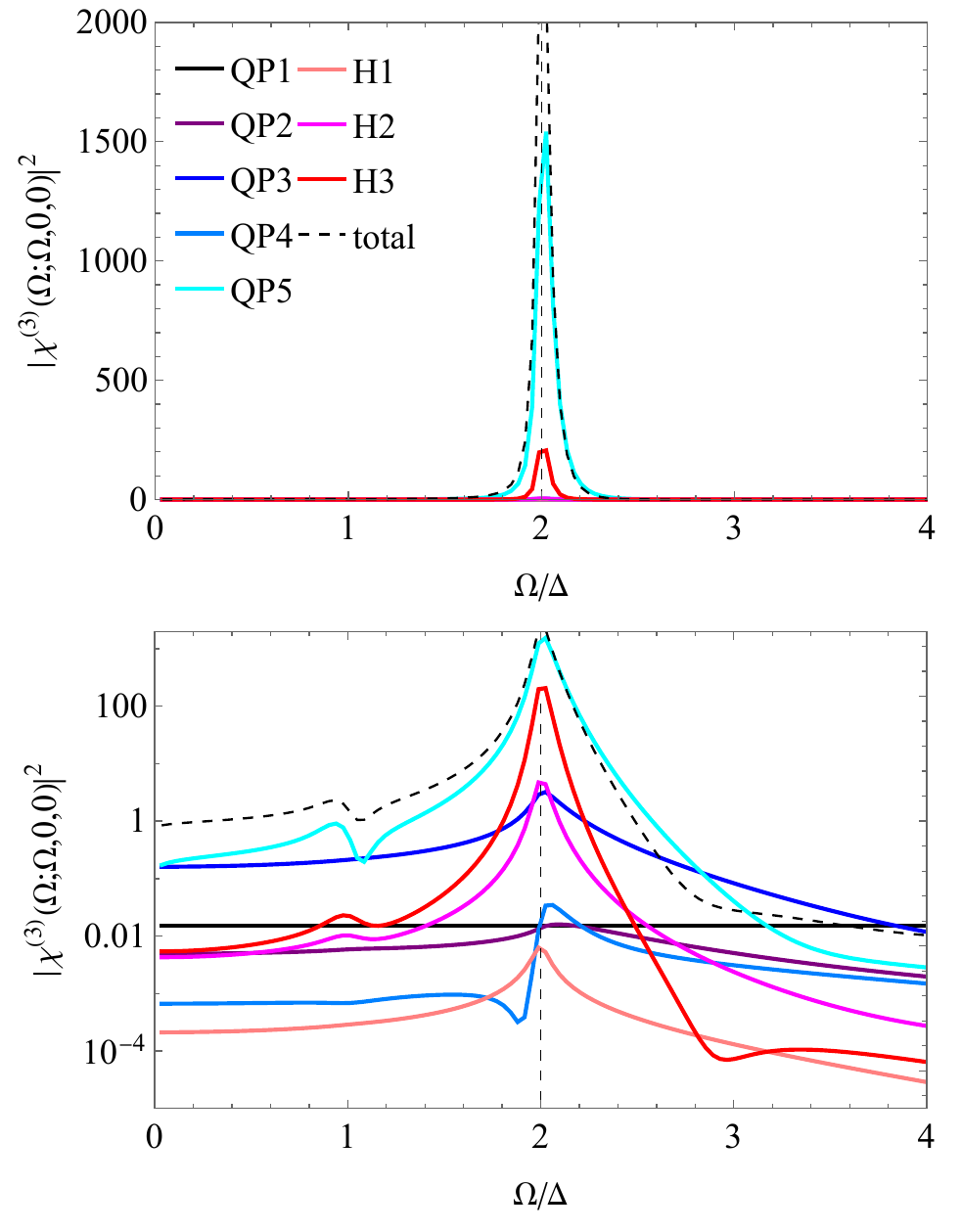}
\caption{Intensity of the dc Kerr susceptibility $\chi^{(3)}(\Omega; \Omega, 0, 0)$ as a function of $\Omega/\Delta$
in the linear scale (top panel) and log scale (bottom)
for the lattice model of disordered superconductors in the clean regime. The parameters are $V=2.5$, $\gamma=0.5$, and $\beta=50$.
The vertical dashed lines indicate $\Omega=2\Delta$.}
\label{fig: dc kerr g05}
\end{figure}

In Fig.~\ref{fig: dc Kerr T}, we plot the temperature dependence of the intensity of the dc Kerr susceptibility $|\chi^{(3)}(\Omega; \Omega, 0, 0)|^2$
at $\Omega=0.1, 0.2, \dots, 0.5$, which shows peaks well below $T_c$.
In contrast to the case of the ac Kerr susceptibility (Fig.~\ref{fig: ac Kerr T}), the peak position clearly depends on $\Omega$.
By comparing with the temperature dependent gap size (Fig.~\ref{fig: gap T}), we can see that the peak position corresponds to
the resonance condition $\Omega=2\Delta$. As we have seen in Fig.~\ref{fig: dc kerr}, the dc Kerr susceptibility $\chi^{(3)}(\Omega; \Omega, 0, 0)$
has the resonance at $\Omega=2\Delta$, which is also observed in the temperature profile.
The result is reminiscent of the experimental observation of the 2DCS signals for NbN superconductors, where
similar resonance structure emerges at $\Omega=2\Delta$. Although these experimental results were obtained for narrow-band pulses, 
they may contain some part of the dc Kerr effect, since the pulses used in the experiment have finite bandwidth and are in between
the narrow-band and broad-band limits. To establish tight connection to experiments, we need to evaluate the full 2DCS signals,
which requires complicated calculations of the integral of the third-order nonlinear susceptibilities with different combination of frequencies
(Eqs.~(\ref{j_NL diagonal}) and (\ref{j_NL horizontal})).

Finally, we investigate the impurity dependence of the ac and dc Kerr susceptibilities. 
As shown for THG, the nonlinear susceptibilities strongly depends on the strength of the impurity scattering $\gamma$.
Depending on the value of $\gamma$, there always arises competition between the Higgs mode and quasiparticle excitations for those nonlinear processes.
Here we take $\gamma=0.5$, which is smaller than that for the previous calculations ($\gamma=2$), and the other parameters remain to be the same 
($V=2.5$ and $\beta=50$ as in Figs.~\ref{fig: ac kerr} and \ref{fig: dc kerr}).
In this parameter set, the gap size is $2\Delta=1.1$ and $\gamma/(2\Delta)=0.45$, which means that the system is in the clean regime.

\begin{table*}[t]
\caption{The relation between the 2DCS in the narrow-band and broad-band limits and third-order nonlinear susceptibilities at selected frequencies.
We restrict ourselves to the case of non-negative frequencies ($\omega_t, \omega_\tau\ge 0$), and decompose the 2DCS signal into those proportional to $A_1^2A_2$
and $A_1A_2^2$ in the broad-band limit.
For each susceptibility, we show characteristic behavior and the dominant physical origin (Higgs mode or quasiparticles) in the clean and dirty regimes.}
\label{table: summary}
\begin{tabular}{ccccccc}
\hline
\hline
pulse & frequency & nonlinear susceptibility & behavior & clean & dirty
\\
\hline
narrow band & $(\omega_t, \omega_\tau)=(3\Omega, \Omega), (3\Omega, 2\Omega)$ & THG: $\chi^{(3)}(3\Omega; \Omega, \Omega, \Omega)$ & resonance at $\Omega=\Delta$ & quasiparticles & Higgs
\\
& $(\omega_t, \omega_\tau)=(\Omega, 0), (\Omega, \Omega), (\Omega, 2\Omega)$ & ac Kerr: $\chi^{(3)}(\Omega; \Omega, \Omega, -\Omega)$ & threshold at $\Omega=2\Delta$ & Higgs & Higgs
\\
& & & & (off-resonant) & (off-resonant)
\\
& & & resonance at $\Omega=\Delta$ & quasiparticles & Higgs
\\
broad band & $(\omega_t, \omega_\tau)=(\Omega, \Omega)$ ($j\propto A_1^2A_2$) & dc Kerr: $\chi^{(3)}(\Omega; \Omega, 0, 0)$ & resonance at $\Omega=2\Delta$ & quasiparticles & Higgs
\\
& $(\omega_t, \omega_\tau)=(\Omega, 0)$ ($j\propto A_1A_2^2$) 
\\
\hline
\hline
\end{tabular}
\end{table*}

In Fig.~\ref{fig: ac kerr g05}, we plot $|\chi^{(3)}(\Omega; \Omega, \Omega, -\Omega)|^2$ for $\gamma=0.5$.
The total contribution is shown by the dashed curve in Fig.~\ref{fig: ac kerr g05}.
As compared to the one for $\gamma=2$ (Fig.~\ref{fig: ac kerr}), the susceptibility still shows the threshold behavior at $\Omega=2\Delta$
dominated by the H3 diagram. The difference is that we can see the decay of the ac Kerr susceptibility at high frequencies
within the plot region. The decay happens roughly in the frequency scale of $\gamma$.
In the clean limit ($\gamma\to 0$), the paramagnetic coupling should disappear, and the contribution of the H3 diagram will vanish.
Thus, we expect that the ac Kerr susceptibility decays more rapidly if we further decrease the value of $\gamma$, and the weight at $\Omega\ge 2\Delta$
is strongly suppressed.
The resonance at $\Omega=\Delta$, on the other hand, is dominated by the quasiparticle contributions (mainly coming from the QP5 and QP3 diagrams).
As in THG, we expect that the two-photon absorption resonance will survive in the clean limit, and is governed by quasiparticle excitations.

In Fig.~\ref{fig: dc kerr g05}, we plot $|\chi^{(3)}(\Omega; \Omega, 0, 0)|^2$ for $\gamma=0.5$. We again find the resonance peak at $\Omega=2\Delta$,
which is dominated by the quasiparticle contributions mainly coming from QP5 diagram. 
We also notice that the weight of the QP3 diagram is enhanced around $\Omega=2\Delta$.
In the clean limit, the paramagnetic coupling to electromagnetic fields disappears, and the resonance at $\Omega=2\Delta$ in the dc Kerr effect
will be governed by the quasiparticles with the diamagnetic coupling.

The results are summarized in Table \ref{table: summary}.
Although the ac Kerr susceptibility at $\Omega\ge 2\Delta$ is dominated by the Higgs mode both in the clean and dirty regimes, 
it only shows the threshold behavior,
and the Higgs mode is far off-resonant. The dc Kerr susceptibility, on the other hand, shows resonance at $\Omega=2\Delta$, which is dominated by the Higgs mode in the dirty regime. 
However, there is a competition between the Higgs mode and quasiparticles depending on the disorder strength as in THG \cite{Jujo2018, MurotaniShimano2019, Silaev2019, TsujiNomura2020, Seibold2021, Derendorf2024}.
In this regard, 2DCS has similar issues on identifying the physical origin of the peak as THG. 
There always exists a possibility that the quasiparticle excitations contribute to the resonance at $\Omega=2\Delta$.
We cannot clearly distinguish the contribution of quasiparticles from that of the Higgs mode only 
by looking at experimental data of the frequency or temperature dependence of 2DCS. 
One needs to compare experimental results with theoretical calculations to understand the origin of the 2DCS signal.
It will also be useful to examine more detailed behavior such as the polarization angle dependence of the pulse field,
as has been the case for THG \cite{Cea2016, Matsunaga2017}.

On the other hand, 2DCS may contain different information than THG.
The 2DCS signal is related to the ac and dc Kerr susceptibilities in the narrow-band and broad-band limits. More generally,
the 2DCS can be described by the integral of the third-order nonlinear susceptibility with different frequency combinations
(Eqs.~(\ref{j_NL diagonal}) and (\ref{j_NL horizontal})). Thus, we can use 2DCS to reconstruct third-order nonlinear susceptibilities
that are otherwise hard to access by other methods.

\section{Summary and outlook}
\label{sec: summary}

To summarize, we have studied the behavior of the two-dimensional coherent spectroscopy (2DCS)
for disordered superconductors in the narrow-band and broad-band limits. We first derive the general formulas (Eqs.~(\ref{j_NL diagonal}) and (\ref{j_NL horizontal}))
that relate the 2DCS signal with the third-order nonlinear susceptibility for general pulse forms (not limited to the narrow-band and broad-band limits).
Then, we apply the formulas to the narrow-band and broad-band limits. In the former, the 2DCS signal is described by the THG and ac Kerr susceptibilities
($\chi^{(3)}(3\Omega; \Omega, \Omega, \Omega)$ and $\chi^{(3)}(\Omega; \Omega, \Omega, -\Omega)$) as in previous studies.
In the latter, we show that the 2DCS signal diverges along the diagonal and horizontal lines in the two-dimensional frequency space.
We decompose the diverging signal into two components according to the pulse amplitude dependence ($A_1^2A_2$ or $A_1A_2^2$).
We find that one of the components is directly proportional to the dc Kerr susceptibility $\chi^{(3)}(\Omega; \Omega, 0, 0)$.

Based on these, we have performed numerical calculations of the ac and dc Kerr susceptibilities for a lattice model of disordered superconductors
within the BCS and self-consistent Born approximations. We demonstrate that the ac Kerr susceptibility shows the threshold behavior
at $\Omega=2\Delta$ rather than resonant enhancement. This is natural since the ac Kerr susceptibility probes the modulation of the linear optical response, 
which shows the threshold at $\Omega=2\Delta$.
The main contribution to the ac Kerr susceptibility comes from the Higgs-mode diagram. However, the Higgs mode mediating
the ac Kerr effect carries zero frequency. Hence the signal only picks up the zero-frequency tail of the Higgs-mode peak (i.e., far off-resonance process).
The dc Kerr susceptibility, on the other hand, shows a clear resonance peak at $\Omega=2\Delta$, which is also dominated by the Higgs mode
in the dirty regime. In this case, the signal arises from the resonant process with the Higgs mode, 
where one in-coming photon carries the fundamental frequency $\Omega$ and another in-coming photon carries zero frequency.
We summarize the results in Table \ref{table: summary}.

We have also studied the temperature dependence of the ac and dc Kerr susceptibilities. In both cases, peaks appear in the temperature profile.
In the case of the ac Kerr susceptibility, the peak position does not depend on the pulse frequency but stays close to the critical temperature.
This reflects the threshold behavior that the spectral weight starts to grow above the gap frequency.
In contrast, the dc Kerr susceptibility exhibits a peak that is always located at $\Omega=2\Delta$.
This mainly arises due to the resonance with the Higgs mode in the dirty regime.
The disorder-strength dependence suggests that there is a competition between the Higgs-mode and quasiparticle contributions
as is the case for THG (see Table \ref{table: summary}). As we approach the clean limit, the quasiparticle excitations give dominant contributions
to the resonance peak of the dc Kerr susceptibility at $\Omega=2\Delta$. For the ac Kerr susceptibility, the Higgs mode continues to contribute
significantly even in the clean regime. We expect that the signal is eventually dominated by quasiparticles in the clean limit, since
the paramagnetic coupling to electromagnetic fields vanishes in that limit and the remaining QP3 diagram has much larger contribution
than the H1 diagram having the diamagnetic coupling.

Let us comment on the relevance of the present results with the experiment \cite{Katsumi2024}. 
As mentioned previously, the experimentally observed peak in
the 2DCS signal at $\Omega=2\Delta$ for NbN superconductors is similar to the behavior of the dc Kerr susceptibility rather than the ac Kerr susceptibility. 
For the latter, we show that the peak in the temperature profile should appear in the vicinity of $T_c$ rather than
the point where $\Omega=2\Delta(T)$ is satisfied.
Thus, the argument based on the ac Kerr susceptibility may not fully explain the experimental results.
We think that in realistic situations the 2DCS signal contains some part of the contribution of the dc Kerr susceptibility
(even though the narrow-band pulses are used in Ref.~\cite{Katsumi2024}), since laser pulses used in experiments have
finite bandwidth, which lies in between the narrow-band and broad-band limits. To fully understand the experimental results,
we need to evaluate the full four-wave mixing susceptibility $\chi^{(3)}(\omega; \omega_1, \omega_2, \omega_3)$ in Eqs.~(\ref{j A_1^2A_2 formula}) and (\ref{j A_1A_2^2 formula}),
which we leave as a future problem.
Let us emphasize that quasiparticles always compete with the Higgs mode in 2DCS (as in THG, see Table \ref{table: summary}).
In particular, the resonance peak in the dc Kerr susceptibility is dominated by quasiparticles in the clean regime.
Thus, the 2DCS alone does not discriminate the Higgs-mode contribution. One has to compare experimental results with 
theoretical calculations to see the physical origin of the 2DCS signal.

Finally, we discuss several open issues related to the present work. First, we have focused on the narrow-band and broad-band limits in 2DCS in this paper.
In the broad-band limit, we further restrict ourselves to the diverging part along the diagonal and horizontal lines in the two-dimensional frequency plane. 
To obtain the full two-dimensional map of the coherent spectroscopy for general pulse wave forms, one has to evaluate the integral of the third-order nonlinear susceptibility
using Eqs.~(\ref{j_NL diagonal}) and (\ref{j_NL horizontal}). While this can be accessible within the present formalism, numerical cost is more demanding.
Second, the effect of strong disorders beyond the self-consistent Born approximation has not been well understood. 
As we have mentioned before, one has to take large system size to keep the scale separation [Eq.~(\ref{scale separation})].
One possible approach is to take account of maximally crossed diagrams (`Cooperon diagrams') for disorders, which are relevant for weak localization effects \cite{Altshuler1980, Rammer}. 
Third, one has to consider multiorbital effects to understand the difference of the experimental observation of 2DCS for NbN and MgB$_2$ \cite{Katsumi2024, Katsumi2025}.
For the latter, there has not been observed a peak in the temperature profile of the 2DCS signal.
Since MgB$_2$ is a typical multi-gap superconductor, the role of another collective mode, the Leggett mode \cite{Leggett1966, Krull2014, Murotani2017, MurotaniShimano2019, Kamatani2022, Nagashima2024, Klein2024, Levitan2024}, in 2DCS should be understood from microscopic calculations. One has to clarify the role of phonon retardation effects as well.
More exotic collective modes in superconductors, such as the Higgs mode with unconventional pairing symmetries \cite{Barlas2013, Schwarz2020, SchwarzManske2020, Vaswani2021, Poniatowski2022} and the Bardasis-Schrieffer mode \cite{Bardasis1961}
are also the target of the future study of 2DCS.

\acknowledgements

We would like to thank N. P. Armitage, K. Katsumi, D. Manske, S. Neri, R. Shimano, and P. Werner for fruitful discussions.
N.T. acknowledges support by JSPS KAKENHI (Grant Nos.~JP24H00191, JP25H01246, and JP25H01251)
and JST FOREST (Grant No.~JPMJFR2131).

\onecolumngrid

\appendix

\section{Keldysh formalism}
\label{app: Green's function}

In this Appendix, we briefly summarize the Keldysh formalism of Green's functions used in the calculation of the nonlinear susceptibilities
for disordered superconductors.
We define the retarded, advanced, and lesser components of the Green's function as follows:
\begin{align}
G_{\bm k}^R(t,t')
&=
-i\theta(t-t')\langle \{ \psi_{\bm k}(t), \psi_{\bm k}^\dagger(t')\}\rangle,
\\
G_{\bm k}^A(t,t')
&=
i\theta(t'-t)\langle \{ \psi_{\bm k}(t), \psi_{\bm k}^\dagger(t')\}\rangle,
\\
G_{\bm k}^<(t,t')
&=
i\langle \psi_{\bm k}^\dagger(t') \psi_{\bm k}(t)\rangle.
\end{align}
Here $\psi_{\bm k}(t)=(c_{\bm k\uparrow}(t), c_{-\bm k\downarrow}^\dagger(t))^T$ is the Nambu spinor,
and each Green's function is in the form of a $2\times 2$ matrix in the Nambu space.
For the lattice model with the Hamiltonian (\ref{Hamiltonian}), the Green's functions are determined 
within the BCS and self-consistent Born approximations as
\begin{align}
G_{\bm k}^{R-1}(\omega)
&=
\omega+i\eta+\mu-\epsilon_{\bm k}\tau_3-\Delta\tau_1-\frac{\gamma^2}{N}\sum_{\bm k} \tau_3 G_{\bm k}^R(\omega) \tau_3,
\label{dyson eq}
\\
G_{\bm k}^A(\omega)
&=
[G_{\bm k}^R(\omega)]^\dagger,
\\
G_{\bm k}^<(\omega)
&=
f(\omega)[G_{\bm k}^A(\omega)-G_{\bm k}^R(\omega)].
\end{align}
In the above, $\tau_i$ ($i=1,2,3$) are Pauli matrices, and $f(\omega)=1/(e^{\beta\omega}+1)$ is the Fermi-Dirac distribution
with inverse temperature $\beta$. Since $G_{\bm k}^R(\omega)$ appears on both sides of Eq.~(\ref{dyson eq}), 
one has to solve it iteratively until convergence is reached. The gap function $\Delta$ is determined from the gap equation (\ref{gap eq}),
which is written in terms of the Green's function as
\begin{align}
\Delta
&=
i\frac{V}{2}\int \frac{d\omega}{2\pi} \frac{1}{N}\sum_{\bm k} {\rm Tr}[\tau_1 G_{\bm k}^<(\omega)].
\label{gap eq G}
\end{align}
We numerically solve Eqs.~(\ref{dyson eq})-(\ref{gap eq G}) self-consistently. 

The optical conductivity can be calculated through Eq.~(\ref{sigma}) with
\begin{align}
\chi^{(1)}(\Omega)
&=
i\int \frac{d\omega}{2\pi} \frac{1}{N} \sum_{\bm k}
{\rm Tr}[\ddot\epsilon_{\bm k}\tau_3 G_{\bm k}(\omega)]^<
+
i\int \frac{d\omega}{2\pi} \frac{1}{N} \sum_{\bm k}
{\rm Tr}[\dot\epsilon_{\bm k} G_{\bm k}(\omega)\dot\epsilon_{\bm k}G_{\bm k}(\omega)]^<.
\end{align}
Here we use the notation of the Langreth rule \cite{Langreth1976, Aoki2014}
(i.e., $[GG]^<=G^R G^<+G^< G^A$),
and $\ddot\epsilon_{\bm k}=\sum_{\mu\nu}\frac{\partial^2\epsilon_{\bm k}}{\partial k_\mu \partial k_\nu} e_\mu e_\nu$
and $\dot\epsilon_{\bm k}=\sum_\mu\frac{\partial\epsilon_{\bm k}}{\partial k_\mu} e_\mu$
are the derivatives of the band dispersion ($e_\mu$ is the polarization vector of driving pulses).
Note that there is no impurity and Higgs-mode vertex correction in $\chi^{(1)}(\Omega)$.

\section{Impurity and Higgs-mode vertices}
\label{app: vertex correction}

In this Appendix, we show the self-consistent equations for the impurity and Higgs-mode vertices that are necessary
in calculating the third-order susceptibility for 2DCS. Those equations are efficiently derived by differentiating the Dyson equation (\ref{dyson eq})
with respect to the driving field as many times as one needs. The derivatives of the self-energy part turn into the vertex corrections.
The derived equations are a simple extension
of those for the third harmonic generation as detailed in Ref.~\cite{TsujiNomura2020}.

The impurity and Higgs-mode vertices are classified according to the type of the coupling to the electric field:
One is the diamagnetic coupling ($\bm A^2$) and the other is the paramagnetic coupling ($\bm j\cdot\bm A$).
When calculating linear and nonlinear susceptibilities within the self-consistent Born approximation, 
one has to include all the noncrossing impurity diagrams that are not contained in the Green's function, 
which forms a series of ladder diagrams.
For the diamagnetic coupling, the impurity vertex $\Lambda_{\alpha\beta}^{\rm dia}(\omega; \Omega)$ satisfies the following equation,
\begin{align}
\Lambda_{\alpha\beta}^{\rm dia}(\omega; \Omega)
&=
\frac{\gamma^2}{N} \sum_{\bm k}
\tau_3 G_{\bm k}(\omega+(\alpha+\beta)\Omega) \ddot\epsilon_{\bm k}\tau_3 G_{\bm k}(\omega) \tau_3
+\frac{\gamma^2}{N} \sum_{\bm k}
\tau_3 G_{\bm k}(\omega) \Lambda_{\alpha\beta}^{\rm dia}(\omega+(\alpha+\beta)\Omega) G_{\bm k}(\omega) \tau_3,
\label{Lambda^dia}
\end{align}
with $\alpha,\beta=1, 0, -1$.
Here the total frequency injected at the diamagnetic vertex is given by $(\alpha+\beta)\Omega$.
We omit the Keldysh labels ($R, A, <$) for simplicity, but it should be understood according to the Langreth rule \cite{Langreth1976, Aoki2014}.
By solving Eq.~(\ref{Lambda^dia}) iteratively, one can include all the impurity ladder corrections to the diamagnetic bare vertex.
Similarly, we have the self-consistent equation for the paramagnetic impurity vertex $\Lambda_{\alpha\beta}^{\rm para}(\omega; \Omega)$,
\begin{align}
\Lambda_{\alpha\beta}^{\rm para}(\omega; \Omega)
&=
\frac{\gamma^2}{N} \sum_{\bm k}
\tau_3 G_{\bm k}(\omega+(\alpha+\beta)\Omega)
(\dot\epsilon_{\bm k} G_{\bm k}(\omega+\alpha\Omega) \dot\epsilon_{\bm k}
+\dot\epsilon_{\bm k} G_{\bm k}(\omega+\beta\Omega) \dot\epsilon_{\bm k})
G_{\bm k}(\omega) \tau_3
\notag
\\
&\quad
+\frac{\gamma^2}{N} \sum_{\bm k}
\tau_3 G_{\bm k}(\omega+(\alpha+\beta)\Omega) \Lambda_{\alpha\beta}^{\rm para}(\omega; \Omega) G_{\bm k}(\omega) \tau_3.
\end{align}
In the case of the paramagnetic coupling, the frequency injected at each 
paramagnetic vertex is given by $\alpha\Omega$ and $\beta\Omega$ ($\alpha, \beta=1, 0, -1$).

The $\tau_1$ vertex (denoted by $\Lambda_{\alpha\beta}^{\tau_1}(\omega; \Omega)$), 
responsible for the Higgs mode propagation, is also subject to the impurity corrections,
satisfying the following equation:
\begin{align}
\Lambda_{\alpha\beta}^{\tau_1}(\omega; \Omega)
&=
\tau_1+\frac{\gamma^2}{N} \sum_{\bm k}
\tau_3 G_{\bm k}(\omega+(\alpha+\beta)\Omega) \Lambda_{\alpha\beta}^{\tau_1}(\omega; \Omega) G_{\bm k}(\omega) \tau_3.
\end{align}
Using the $\tau_1$ vertex and impurity vertex, 
we can write down the self-consistent equation for the second derivative of the gap function with respect to the vector potential
in the diamagnetic channel as
\begin{align}
\ddot\Delta_{\alpha\beta}^{\rm dia}(\Omega)
&=
\frac{i}{2} \frac{V}{N} \sum_{\bm k} \int \frac{d\omega}{2\pi} {\rm Tr}[\tau_1
G_{\bm k}(\omega+(\alpha+\beta)\Omega)(\ddot\epsilon_{\bm k}\tau_3+\Lambda_{\alpha\beta}^{\rm dia}(\omega; \Omega)
+\Lambda_{\alpha\beta}^{\tau_1}(\omega; \Omega)\ddot\Delta_{\alpha\beta}^{\rm dia}(\Omega))
G_{\bm k}(\omega)]^<,
\end{align}
and in the paramagnetic channel as
\begin{align}
\ddot\Delta_{\alpha\beta}^{\rm para}(\Omega)
&=
\frac{i}{2} \frac{V}{N} \sum_{\bm k} \int \frac{d\omega}{2\pi} {\rm Tr}[\tau_1
G_{\bm k}(\omega+(\alpha+\beta)\Omega)(\dot\epsilon_{\bm k}G_{\bm k}(\omega+\alpha\Omega)\dot\epsilon_{\bm k}
+\dot\epsilon_{\bm k}G_{\bm k}(\omega+\beta\Omega)\dot\epsilon_{\bm k}
+\Lambda_{\alpha\beta}^{\rm para}(\omega; \Omega)
\notag
\\
&\quad
+\Lambda_{\alpha\beta}^{\tau_1}(\omega; \Omega)\ddot\Delta_{\alpha\beta}^{\rm para}(\Omega))
G_{\bm k}(\omega)]^<.
\end{align}
With the $\tau_1$ vertex, the Higgs-mode propagator can be written as
\begin{align}
H(2\Omega)
&=
\left[
1-\frac{i}{2} \frac{V}{N} \sum_{\bm k} \int \frac{d\omega}{2\pi} {\rm Tr}[\tau_1
G_{\bm k}(\omega+2\Omega)
\Lambda_{++}^{\tau_1}(\omega; \Omega)
G_{\bm k}(\omega)]^<
\right]^{-1}V.
\end{align}

\section{Third-order nonlinear susceptibility for 2DCS}
\label{app: chi3}


In this Appendix, we give the explicit form of the third-order nonlinear susceptibility for 2DCS.
The corresponding diagrams are shown in Fig.~\ref{fig: kerr diagram full} in the main text.
For each diagram, we write down the explicit expression of the ac Kerr susceptibility using the Green's function
and impurity and Higgs-mode vertex corrections below:
\begin{align}
\chi_{\rm QP1}^{(3)}(\Omega; +\Omega, +\Omega, -\Omega)
&=
\frac{i}{6} \int \frac{d\omega}{2\pi} \frac{1}{N} \sum_{\bm k}
{\rm Tr}[\ddddot\epsilon_{\bm k}\tau_3 G_{\bm k}(\omega)]^<,
\\
\chi_{\rm QP2}^{(3)}(\Omega; +\Omega, +\Omega, -\Omega)
&=
\frac{i}{3} \int \frac{d\omega}{2\pi} \frac{1}{N} \sum_{\bm k}
{\rm Tr}[\dddot\epsilon_{\bm k} G_{\bm k}(\omega+\Omega) \dot\epsilon_{\bm k} G_{\bm k}(\omega)]^<
\notag
\\
&\quad
+\frac{i}{6} \int \frac{d\omega}{2\pi} \frac{1}{N} \sum_{\bm k}
{\rm Tr}[\dddot\epsilon_{\bm k} G_{\bm k}(\omega-\Omega) \dot\epsilon_{\bm k} G_{\bm k}(\omega)]^<
\notag
\\
&\quad
+\frac{i}{6} \int \frac{d\omega}{2\pi} \frac{1}{N} \sum_{\bm k}
{\rm Tr}[\dot\epsilon_{\bm k}G_{\bm k}(\omega+\Omega)\dddot\epsilon_{\bm k}G_{\bm k}(\omega)]^<,
\end{align}

\begin{align}
\chi_{\rm QP3}^{(3)}(\Omega; +\Omega, +\Omega, -\Omega)
&=
\frac{i}{6} \int \frac{d\omega}{2\pi} \frac{1}{N} \sum_{\bm k}
{\rm Tr}[
\ddot\epsilon_{\bm k}\tau_3 G_{\bm k}(\omega+2\Omega) \ddot\epsilon_{\bm k} \tau_3 G_{\bm k}(\omega)
]^<
\notag
\\
&\quad
+\frac{i}{3} \int \frac{d\omega}{2\pi} \frac{1}{N} \sum_{\bm k}
{\rm Tr}[
\ddot\epsilon_{\bm k}\tau_3 G_{\bm k}(\omega) \ddot\epsilon_{\bm k} \tau_3 G_{\bm k}(\omega)
]^<
\notag
\\
&\quad
+\frac{i}{6} \int \frac{d\omega}{2\pi} \frac{1}{N} \sum_{\bm k}
{\rm Tr}[
\ddot\epsilon_{\bm k}\tau_3 G_{\bm k}(\omega+2\Omega) \Lambda_{1,1}^{\rm dia}(\omega; \Omega) G_{\bm k}(\omega)
]^<
\notag
\\
&\quad
+\frac{i}{3} \int \frac{d\omega}{2\pi} \frac{1}{N} \sum_{\bm k}
{\rm Tr}[
\ddot\epsilon_{\bm k}\tau_3 G_{\bm k}(\omega) \Lambda_{1,-1}^{\rm dia}(\omega; \Omega) G_{\bm k}(\omega)
]^<,
\\
\chi_{\rm QP4}^{(3)}(\Omega; +\Omega, +\Omega, -\Omega)
&=
\frac{i}{3} \int \frac{d\omega}{2\pi} \frac{1}{N} \sum_{\bm k}
{\rm Tr}[
\ddot\epsilon_{\bm k}\tau_3 G_{\bm k}(\omega+2\Omega) \dot\epsilon_{\bm k} G_{\bm k}(\omega+\Omega) \dot\epsilon_{\bm k} G_{\bm k}(\omega)
]^<
\notag
\\
&\quad
+\frac{i}{3} \int \frac{d\omega}{2\pi} \frac{1}{N} \sum_{\bm k}
{\rm Tr}[
\ddot\epsilon_{\bm k}\tau_3 G_{\bm k}(\omega) \dot\epsilon_{\bm k} G_{\bm k}(\omega+\Omega) \dot\epsilon_{\bm k} G_{\bm k}(\omega)
]^<
\notag
\\
&\quad
+\frac{i}{3} \int \frac{d\omega}{2\pi} \frac{1}{N} \sum_{\bm k}
{\rm Tr}[
\ddot\epsilon_{\bm k}\tau_3 G_{\bm k}(\omega) \dot\epsilon_{\bm k} G_{\bm k}(\omega-\Omega) \dot\epsilon_{\bm k} G_{\bm k}(\omega)
]^<
\notag
\\
&\quad
+\frac{i}{6} \int \frac{d\omega}{2\pi} \frac{1}{N} \sum_{\bm k}
{\rm Tr}[
\dot\epsilon_{\bm k} G_{\bm k}(\omega+\Omega) \dot\epsilon_{\bm k} G_{\bm k}(\omega+2\Omega) \ddot\epsilon_{\bm k}\tau_3 G_{\bm k}(\omega)
]^<
\notag
\\
&\quad
+\frac{i}{3} \int \frac{d\omega}{2\pi} \frac{1}{N} \sum_{\bm k}
{\rm Tr}[
\dot\epsilon_{\bm k} G_{\bm k}(\omega+\Omega) \dot\epsilon_{\bm k} G_{\bm k}(\omega) \ddot\epsilon_{\bm k}\tau_3 G_{\bm k}(\omega)
]^<
\notag
\\
&\quad
+\frac{i}{6} \int \frac{d\omega}{2\pi} \frac{1}{N} \sum_{\bm k}
{\rm Tr}[
\dot\epsilon_{\bm k} G_{\bm k}(\omega+\Omega) \ddot\epsilon_{\bm k}\tau_3 G_{\bm k}(\omega-\Omega) \dot\epsilon_{\bm k} G_{\bm k}(\omega)
]^<
\notag
\\
&\quad
+\frac{i}{3} \int \frac{d\omega}{2\pi} \frac{1}{N} \sum_{\bm k}
{\rm Tr}[
\dot\epsilon_{\bm k} G_{\bm k}(\omega+\Omega) \ddot\epsilon_{\bm k}\tau_3 G_{\bm k}(\omega+\Omega) \dot\epsilon_{\bm k} G_{\bm k}(\omega)
]^<
\notag
\\
&\quad
+\frac{i}{6} \int \frac{d\omega}{2\pi} \frac{1}{N} \sum_{\bm k}
{\rm Tr}[
\ddot\epsilon_{\bm k}\tau_3 G_{\bm k}(\omega+2\Omega) \Lambda_{1,1}^{\rm para}(\omega; \Omega) G_{\bm k}(\omega)
]^<
\notag
\\
&\quad
+\frac{i}{3} \int \frac{d\omega}{2\pi} \frac{1}{N} \sum_{\bm k}
{\rm Tr}[
\ddot\epsilon_{\bm k}\tau_3 G_{\bm k}(\omega) \Lambda_{1,-1}^{\rm para}(\omega; \Omega) G_{\bm k}(\omega)
]^<
\notag
\\
&\quad
+\frac{i}{6} \int \frac{d\omega}{2\pi} \frac{1}{N} \sum_{\bm k}
{\rm Tr}[
\dot\epsilon_{\bm k} G_{\bm k}(\omega+\Omega) \dot\epsilon_{\bm k} G_{\bm k}(\omega+2\Omega) \Lambda_{1,1}^{\rm dia}(\omega; \Omega) G_{\bm k}(\omega)
]^<
\notag
\\
&\quad
+\frac{i}{3} \int \frac{d\omega}{2\pi} \frac{1}{N} \sum_{\bm k}
{\rm Tr}[
\dot\epsilon_{\bm k} G_{\bm k}(\omega+\Omega) \dot\epsilon_{\bm k} G_{\bm k}(\omega) \Lambda_{1,-1}^{\rm dia}(\omega; \Omega) G_{\bm k}(\omega)
]^<
\notag
\\
&\quad
+\frac{i}{6} \int \frac{d\omega}{2\pi} \frac{1}{N} \sum_{\bm k}
{\rm Tr}[
\dot\epsilon_{\bm k} G_{\bm k}(\omega+\Omega) \Lambda_{1,1}^{\rm dia}(\omega-\Omega; \Omega) G_{\bm k}(\omega-\Omega) \dot\epsilon_{\bm k} G_{\bm k}(\omega)
]^<
\notag
\\
&\quad
+\frac{i}{3} \int \frac{d\omega}{2\pi} \frac{1}{N} \sum_{\bm k}
{\rm Tr}[
\dot\epsilon_{\bm k} G_{\bm k}(\omega+\Omega) \Lambda_{1,-1}^{\rm dia}(\omega+\Omega; \Omega) G_{\bm k}(\omega+\Omega) \dot\epsilon_{\bm k} G_{\bm k}(\omega)
]^<,
\end{align}

\begin{align}
\chi_{\rm QP5}^{(3)}(\Omega; +\Omega, +\Omega, -\Omega)
&=
\frac{i}{3} \int \frac{d\omega}{2\pi} \frac{1}{N} \sum_{\bm k}
{\rm Tr}[
G_{\bm k}(\omega+\Omega) \dot\epsilon_{\bm k} G_{\bm k}(\omega+2\Omega) \dot\epsilon_{\bm k} G_{\bm k}(\omega+\Omega) \dot\epsilon_{\bm k} G_{\bm k}(\omega)
]^<
\notag
\\
&\quad
+\frac{i}{3} \int \frac{d\omega}{2\pi} \frac{1}{N} \sum_{\bm k}
{\rm Tr}[
G_{\bm k}(\omega+\Omega) \dot\epsilon_{\bm k} G_{\bm k}(\omega) \dot\epsilon_{\bm k} G_{\bm k}(\omega+\Omega) \dot\epsilon_{\bm k} G_{\bm k}(\omega)
]^<
\notag
\\
&\quad
+\frac{i}{3} \int \frac{d\omega}{2\pi} \frac{1}{N} \sum_{\bm k}
{\rm Tr}[
G_{\bm k}(\omega+\Omega) \dot\epsilon_{\bm k} G_{\bm k}(\omega) \dot\epsilon_{\bm k} G_{\bm k}(\omega-\Omega) \dot\epsilon_{\bm k} G_{\bm k}(\omega)
]^<
\notag
\\
&\quad
+\frac{i}{6} \int \frac{d\omega}{2\pi} \frac{1}{N} \sum_{\bm k}
{\rm Tr}[
G_{\bm k}(\omega+\Omega) \dot\epsilon_{\bm k} G_{\bm k}(\omega+2\Omega) \Lambda_{1,1}^{\rm para}(\omega; \Omega) G_{\bm k}(\omega)
]^<
\notag
\\
&\quad
+\frac{i}{3} \int \frac{d\omega}{2\pi} \frac{1}{N} \sum_{\bm k}
{\rm Tr}[
G_{\bm k}(\omega+\Omega) \dot\epsilon_{\bm k} G_{\bm k}(\omega) \Lambda_{1,-1}^{\rm para}(\omega; \Omega) G_{\bm k}(\omega)
]^<
\notag
\\
&\quad
+\frac{i}{6} \int \frac{d\omega}{2\pi} \frac{1}{N} \sum_{\bm k}
{\rm Tr}[
G_{\bm k}(\omega+\Omega) \Lambda_{1,1}^{\rm para}(\omega-\Omega; \Omega) G_{\bm k}(\omega-\Omega) \dot\epsilon_{\bm k} G_{\bm k}(\omega)
]^<
\notag
\\
&\quad
+\frac{i}{3} \int \frac{d\omega}{2\pi} \frac{1}{N} \sum_{\bm k}
{\rm Tr}[
G_{\bm k}(\omega+\Omega) \Lambda_{1,-1}^{\rm para}(\omega+\Omega; \Omega) G_{\bm k}(\omega+\Omega) \dot\epsilon_{\bm k} G_{\bm k}(\omega)
]^<,
\\
\chi_{\rm H1}^{(3)}(\Omega; +\Omega, +\Omega, -\Omega)
&=
\frac{i}{6} \int \frac{d\omega}{2\pi} \frac{1}{N} \sum_{\bm k}
{\rm Tr}[
\ddot\epsilon_{\bm k}\tau_3 G_{\bm k}(\omega+2\Omega) \ddot\Delta_{1,1}^{\rm dia}(\Omega)\Lambda_{1,1}^{\tau_1}(\omega; \Omega) G_{\bm k}(\omega)
]^<
\notag
\\
&\quad
+\frac{i}{3} \int \frac{d\omega}{2\pi} \frac{1}{N} \sum_{\bm k}
{\rm Tr}[
\ddot\epsilon_{\bm k}\tau_3 G_{\bm k}(\omega) \ddot\Delta_{1,-1}^{\rm dia}(\Omega)\Lambda_{1,-1}^{\tau_1}(\omega; \Omega) G_{\bm k}(\omega)
]^<,
\\
\chi_{\rm H2}^{(3)}(\Omega; +\Omega, +\Omega, -\Omega)
&=
\frac{i}{6} \int \frac{d\omega}{2\pi} \frac{1}{N} \sum_{\bm k}
{\rm Tr}[
\ddot\epsilon_{\bm k}\tau_3 G_{\bm k}(\omega+2\Omega) \ddot\Delta_{1,1}^{\rm para}(\Omega)\Lambda_{1,1}^{\tau_1}(\omega; \Omega) G_{\bm k}(\omega)
]^<
\notag
\\
&\quad
+\frac{i}{3} \int \frac{d\omega}{2\pi} \frac{1}{N} \sum_{\bm k}
{\rm Tr}[
\ddot\epsilon_{\bm k}\tau_3 G_{\bm k}(\omega) \ddot\Delta_{1,-1}^{\rm para}(\Omega)\Lambda_{1,-1}^{\tau_1}(\omega; \Omega) G_{\bm k}(\omega)
]^<
\notag
\\
&\quad
+\frac{i}{6} \int \frac{d\omega}{2\pi} \frac{1}{N} \sum_{\bm k}
{\rm Tr}[
\dot\epsilon_{\bm k} G_{\bm k}(\omega+\Omega) \dot\epsilon_{\bm k} G_{\bm k}(\omega+2\Omega) \ddot\Delta_{1,1}^{\rm dia}(\Omega)\Lambda_{1,1}^{\tau_1}(\omega; \Omega) G_{\bm k}(\omega)
]^<
\notag
\\
&\quad
+\frac{i}{3} \int \frac{d\omega}{2\pi} \frac{1}{N} \sum_{\bm k}
{\rm Tr}[
\dot\epsilon_{\bm k} G_{\bm k}(\omega+\Omega) \dot\epsilon_{\bm k} G_{\bm k}(\omega) \ddot\Delta_{1,-1}^{\rm dia}(\Omega)\Lambda_{1,-1}^{\tau_1}(\omega; \Omega) G_{\bm k}(\omega)
]^<
\notag
\\
&\quad
+\frac{i}{6} \int \frac{d\omega}{2\pi} \frac{1}{N} \sum_{\bm k}
{\rm Tr}[
\dot\epsilon_{\bm k} G_{\bm k}(\omega+\Omega) \ddot\Delta_{1,1}^{\rm dia}(\Omega)\Lambda_{1,1}^{\tau_1}(\omega-\Omega; \Omega) G_{\bm k}(\omega-\Omega) \dot\epsilon_{\bm k} G_{\bm k}(\omega)
]^<
\notag
\\
&\quad
+\frac{i}{3} \int \frac{d\omega}{2\pi} \frac{1}{N} \sum_{\bm k}
{\rm Tr}[
\dot\epsilon_{\bm k} G_{\bm k}(\omega+\Omega) \ddot\Delta_{1,-1}^{\rm dia}(\Omega)\Lambda_{1,-1}^{\tau_1}(\omega+\Omega; \Omega) G_{\bm k}(\omega+\Omega) \dot\epsilon_{\bm k} G_{\bm k}(\omega)
]^<,
\\
\chi_{\rm H3}^{(3)}(\Omega; +\Omega, +\Omega, -\Omega)
&=
\frac{i}{6} \int \frac{d\omega}{2\pi} \frac{1}{N} \sum_{\bm k}
{\rm Tr}[
G_{\bm k}(\omega+\Omega) \dot\epsilon_{\bm k} G_{\bm k}(\omega+2\Omega) \ddot\Delta_{1,1}^{\rm para}(\Omega)\Lambda_{1,1}^{\tau_1}(\omega; \Omega) G_{\bm k}(\omega)
]^<
\notag
\\
&\quad
+\frac{i}{3} \int \frac{d\omega}{2\pi} \frac{1}{N} \sum_{\bm k}
{\rm Tr}[
G_{\bm k}(\omega+\Omega) \dot\epsilon_{\bm k} G_{\bm k}(\omega) \ddot\Delta_{1,-1}^{\rm para}(\Omega)\Lambda_{1,-1}^{\tau_1}(\omega; \Omega) G_{\bm k}(\omega)
]^<
\notag
\\
&\quad
+\frac{i}{6} \int \frac{d\omega}{2\pi} \frac{1}{N} \sum_{\bm k}
{\rm Tr}[
G_{\bm k}(\omega+\Omega) \ddot\Delta_{1,1}^{\rm para}(\Omega)\Lambda_{1,1}^{\tau_1}(\omega-\Omega; \Omega) G_{\bm k}(\omega-\Omega) \dot\epsilon_{\bm k} G_{\bm k}(\omega)
]^<
\notag
\\
&\quad
+\frac{i}{3} \int \frac{d\omega}{2\pi} \frac{1}{N} \sum_{\bm k}
{\rm Tr}[
G_{\bm k}(\omega+\Omega) \ddot\Delta_{1,-1}^{\rm para}(\Omega)\Lambda_{1,-1}^{\tau_1}(\omega+\Omega; \Omega) G_{\bm k}(\omega+\Omega) \dot\epsilon_{\bm k} G_{\bm k}(\omega)
]^<.
\end{align}
Here $\dddot\epsilon_{\bm k}=\sum_{\mu\nu\lambda}\frac{\partial^3\epsilon_{\bm k}}{\partial k_\mu\partial k_\nu\partial k_\lambda}
e_\mu e_\nu e_\lambda$
and $\ddddot\epsilon_{\bm k}=\sum_{\mu\nu\lambda}\frac{\partial^4\epsilon_{\bm k}}{\partial k_\mu\partial k_\nu\partial k_\lambda\partial k_\kappa}
e_\mu e_\nu e_\lambda e_\kappa$.

Similar expressions for the dc Kerr susceptibility can be found below:
\begin{align}
\chi_{\rm QP1}^{(3)}(\Omega; \Omega, 0, 0)
&=
\frac{i}{6} \int \frac{d\omega}{2\pi} \frac{1}{N} \sum_{\bm k}
{\rm Tr}[\ddddot\epsilon_{\bm k}\tau_3 G_{\bm k}(\omega)]^<,
\\
\chi_{\rm QP2}^{(3)}(\Omega; \Omega, 0, 0)
&=
\frac{i}{3} \int \frac{d\omega}{2\pi} \frac{1}{N} \sum_{\bm k}
{\rm Tr}[\dddot\epsilon_{\bm k} G_{\bm k}(\omega) \dot\epsilon_{\bm k} G_{\bm k}(\omega)]^<
\notag
\\
&\quad
+\frac{i}{6} \int \frac{d\omega}{2\pi} \frac{1}{N} \sum_{\bm k}
{\rm Tr}[\dddot\epsilon_{\bm k} G_{\bm k}(\omega+\Omega) \dot\epsilon_{\bm k} G_{\bm k}(\omega)]^<
\notag
\\
&\quad
+\frac{i}{6} \int \frac{d\omega}{2\pi} \frac{1}{N} \sum_{\bm k}
{\rm Tr}[\dot\epsilon_{\bm k}G_{\bm k}(\omega+\Omega)\dddot\epsilon_{\bm k}G_{\bm k}(\omega)]^<,
\\
\chi_{\rm QP3}^{(3)}(\Omega; \Omega, 0, 0)
&=
\frac{i}{6} \int \frac{d\omega}{2\pi} \frac{1}{N} \sum_{\bm k}
{\rm Tr}[
\ddot\epsilon_{\bm k}\tau_3 G_{\bm k}(\omega) \ddot\epsilon_{\bm k} \tau_3 G_{\bm k}(\omega)
]^<
\notag
\\
&\quad
+\frac{i}{3} \int \frac{d\omega}{2\pi} \frac{1}{N} \sum_{\bm k}
{\rm Tr}[
\ddot\epsilon_{\bm k}\tau_3 G_{\bm k}(\omega+\Omega) \ddot\epsilon_{\bm k} \tau_3 G_{\bm k}(\omega)
]^<
\notag
\\
&\quad
+\frac{i}{6} \int \frac{d\omega}{2\pi} \frac{1}{N} \sum_{\bm k}
{\rm Tr}[
\ddot\epsilon_{\bm k}\tau_3 G_{\bm k}(\omega) \Lambda_{0,0}^{\rm dia}(\omega; \Omega) G_{\bm k}(\omega)
]^<
\notag
\\
&\quad
+\frac{i}{3} \int \frac{d\omega}{2\pi} \frac{1}{N} \sum_{\bm k}
{\rm Tr}[
\ddot\epsilon_{\bm k}\tau_3 G_{\bm k}(\omega+\Omega) \Lambda_{1,0}^{\rm dia}(\omega; \Omega) G_{\bm k}(\omega)
]^<,
\\
\chi_{\rm QP4}^{(3)}(\Omega; \Omega, 0, 0)
&=
\frac{i}{3} \int \frac{d\omega}{2\pi} \frac{1}{N} \sum_{\bm k}
{\rm Tr}[
\ddot\epsilon_{\bm k}\tau_3 G_{\bm k}(\omega+\Omega) \dot\epsilon_{\bm k} G_{\bm k}(\omega+\Omega) \dot\epsilon_{\bm k} G_{\bm k}(\omega)
]^<
\notag
\\
&\quad
+\frac{i}{3} \int \frac{d\omega}{2\pi} \frac{1}{N} \sum_{\bm k}
{\rm Tr}[
\ddot\epsilon_{\bm k}\tau_3 G_{\bm k}(\omega+\Omega) \dot\epsilon_{\bm k} G_{\bm k}(\omega) \dot\epsilon_{\bm k} G_{\bm k}(\omega)
]^<
\notag
\\
&\quad
+\frac{i}{3} \int \frac{d\omega}{2\pi} \frac{1}{N} \sum_{\bm k}
{\rm Tr}[
\ddot\epsilon_{\bm k}\tau_3 G_{\bm k}(\omega) \dot\epsilon_{\bm k} G_{\bm k}(\omega) \dot\epsilon_{\bm k} G_{\bm k}(\omega)
]^<
\notag
\\
&\quad
+\frac{i}{6} \int \frac{d\omega}{2\pi} \frac{1}{N} \sum_{\bm k}
{\rm Tr}[
\dot\epsilon_{\bm k} G_{\bm k}(\omega+\Omega) \dot\epsilon_{\bm k} G_{\bm k}(\omega) \ddot\epsilon_{\bm k}\tau_3 G_{\bm k}(\omega)
]^<
\notag
\\
&\quad
+\frac{i}{3} \int \frac{d\omega}{2\pi} \frac{1}{N} \sum_{\bm k}
{\rm Tr}[
\dot\epsilon_{\bm k} G_{\bm k}(\omega+\Omega) \dot\epsilon_{\bm k} G_{\bm k}(\omega+\Omega) \ddot\epsilon_{\bm k}\tau_3 G_{\bm k}(\omega)
]^<
\notag
\\
&\quad
+\frac{i}{6} \int \frac{d\omega}{2\pi} \frac{1}{N} \sum_{\bm k}
{\rm Tr}[
\dot\epsilon_{\bm k} G_{\bm k}(\omega+\Omega) \ddot\epsilon_{\bm k}\tau_3 G_{\bm k}(\omega+\Omega) \dot\epsilon_{\bm k} G_{\bm k}(\omega)
]^<
\notag
\\
&\quad
+\frac{i}{3} \int \frac{d\omega}{2\pi} \frac{1}{N} \sum_{\bm k}
{\rm Tr}[
\dot\epsilon_{\bm k} G_{\bm k}(\omega+\Omega) \ddot\epsilon_{\bm k}\tau_3 G_{\bm k}(\omega) \dot\epsilon_{\bm k} G_{\bm k}(\omega)
]^<
\notag
\\
&\quad
+\frac{i}{6} \int \frac{d\omega}{2\pi} \frac{1}{N} \sum_{\bm k}
{\rm Tr}[
\ddot\epsilon_{\bm k}\tau_3 G_{\bm k}(\omega) \Lambda_{0,0}^{\rm para}(\omega; \Omega) G_{\bm k}(\omega)
]^<
\notag
\\
&\quad
+\frac{i}{3} \int \frac{d\omega}{2\pi} \frac{1}{N} \sum_{\bm k}
{\rm Tr}[
\ddot\epsilon_{\bm k}\tau_3 G_{\bm k}(\omega+\Omega) \Lambda_{1,0}^{\rm para}(\omega; \Omega) G_{\bm k}(\omega)
]^<
\notag
\\
&\quad
+\frac{i}{6} \int \frac{d\omega}{2\pi} \frac{1}{N} \sum_{\bm k}
{\rm Tr}[
\dot\epsilon_{\bm k} G_{\bm k}(\omega+\Omega) \dot\epsilon_{\bm k} G_{\bm k}(\omega) \Lambda_{0,0}^{\rm dia}(\omega; \Omega) G_{\bm k}(\omega)
]^<
\notag
\\
&\quad
+\frac{i}{3} \int \frac{d\omega}{2\pi} \frac{1}{N} \sum_{\bm k}
{\rm Tr}[
\dot\epsilon_{\bm k} G_{\bm k}(\omega+\Omega) \dot\epsilon_{\bm k} G_{\bm k}(\omega+\Omega) \Lambda_{1,0}^{\rm dia}(\omega; \Omega) G_{\bm k}(\omega)
]^<
\notag
\\
&\quad
+\frac{i}{6} \int \frac{d\omega}{2\pi} \frac{1}{N} \sum_{\bm k}
{\rm Tr}[
\dot\epsilon_{\bm k} G_{\bm k}(\omega+\Omega) \Lambda_{0,0}^{\rm dia}(\omega+\Omega; \Omega) G_{\bm k}(\omega+\Omega) \dot\epsilon_{\bm k} G_{\bm k}(\omega)
]^<
\notag
\\
&\quad
+\frac{i}{3} \int \frac{d\omega}{2\pi} \frac{1}{N} \sum_{\bm k}
{\rm Tr}[
\dot\epsilon_{\bm k} G_{\bm k}(\omega+\Omega) \Lambda_{1,0}^{\rm dia}(\omega; \Omega) G_{\bm k}(\omega) \dot\epsilon_{\bm k} G_{\bm k}(\omega)
]^<,
\end{align}

\begin{align}
\chi_{\rm QP5}^{(3)}(\Omega; \Omega, 0, 0)
&=
\frac{i}{3} \int \frac{d\omega}{2\pi} \frac{1}{N} \sum_{\bm k}
{\rm Tr}[
G_{\bm k}(\omega+\Omega) \dot\epsilon_{\bm k} G_{\bm k}(\omega) \dot\epsilon_{\bm k} G_{\bm k}(\omega) \dot\epsilon_{\bm k} G_{\bm k}(\omega)
]^<
\notag
\\
&\quad
+\frac{i}{3} \int \frac{d\omega}{2\pi} \frac{1}{N} \sum_{\bm k}
{\rm Tr}[
G_{\bm k}(\omega+\Omega) \dot\epsilon_{\bm k} G_{\bm k}(\omega+\Omega) \dot\epsilon_{\bm k} G_{\bm k}(\omega) \dot\epsilon_{\bm k} G_{\bm k}(\omega)
]^<
\notag
\\
&\quad
+\frac{i}{3} \int \frac{d\omega}{2\pi} \frac{1}{N} \sum_{\bm k}
{\rm Tr}[
G_{\bm k}(\omega+\Omega) \dot\epsilon_{\bm k} G_{\bm k}(\omega+\Omega) \dot\epsilon_{\bm k} G_{\bm k}(\omega+\Omega) \dot\epsilon_{\bm k} G_{\bm k}(\omega)
]^<
\notag
\\
&\quad
+\frac{i}{6} \int \frac{d\omega}{2\pi} \frac{1}{N} \sum_{\bm k}
{\rm Tr}[
G_{\bm k}(\omega+\Omega) \dot\epsilon_{\bm k} G_{\bm k}(\omega) \Lambda_{0,0}^{\rm para}(\omega; \Omega) G_{\bm k}(\omega)
]^<
\notag
\\
&\quad
+\frac{i}{3} \int \frac{d\omega}{2\pi} \frac{1}{N} \sum_{\bm k}
{\rm Tr}[
G_{\bm k}(\omega+\Omega) \dot\epsilon_{\bm k} G_{\bm k}(\omega+\Omega) \Lambda_{1,0}^{\rm para}(\omega; \Omega) G_{\bm k}(\omega)
]^<
\notag
\\
&\quad
+\frac{i}{6} \int \frac{d\omega}{2\pi} \frac{1}{N} \sum_{\bm k}
{\rm Tr}[
G_{\bm k}(\omega+\Omega) \Lambda_{0,0}^{\rm para}(\omega+\Omega; \Omega) G_{\bm k}(\omega+\Omega) \dot\epsilon_{\bm k} G_{\bm k}(\omega)
]^<
\notag
\\
&\quad
+\frac{i}{3} \int \frac{d\omega}{2\pi} \frac{1}{N} \sum_{\bm k}
{\rm Tr}[
G_{\bm k}(\omega+\Omega) \Lambda_{1,0}^{\rm para}(\omega; \Omega) G_{\bm k}(\omega) \dot\epsilon_{\bm k} G_{\bm k}(\omega)
]^<,
\\
\chi_{\rm H1}^{(3)}(\Omega; \Omega, 0, 0)
&=
\frac{i}{6} \int \frac{d\omega}{2\pi} \frac{1}{N} \sum_{\bm k}
{\rm Tr}[
\ddot\epsilon_{\bm k}\tau_3 G_{\bm k}(\omega) \ddot\Delta_{0,0}^{\rm dia}(\Omega)\Lambda_{0,0}^{\tau_1}(\omega; \Omega) G_{\bm k}(\omega)
]^<
\notag
\\
&\quad
+\frac{i}{3} \int \frac{d\omega}{2\pi} \frac{1}{N} \sum_{\bm k}
{\rm Tr}[
\ddot\epsilon_{\bm k}\tau_3 G_{\bm k}(\omega+\Omega) \ddot\Delta_{1,0}^{\rm dia}(\Omega)\Lambda_{1,0}^{\tau_1}(\omega; \Omega) G_{\bm k}(\omega)
]^<,
\\
\chi_{\rm H2}^{(3)}(\Omega; \Omega, 0, 0)
&=
\frac{i}{6} \int \frac{d\omega}{2\pi} \frac{1}{N} \sum_{\bm k}
{\rm Tr}[
\ddot\epsilon_{\bm k}\tau_3 G_{\bm k}(\omega) \ddot\Delta_{0,0}^{\rm para}(\Omega)\Lambda_{0,0}^{\tau_1}(\omega; \Omega) G_{\bm k}(\omega)
]^<
\notag
\\
&\quad
+\frac{i}{3} \int \frac{d\omega}{2\pi} \frac{1}{N} \sum_{\bm k}
{\rm Tr}[
\ddot\epsilon_{\bm k}\tau_3 G_{\bm k}(\omega+\Omega) \ddot\Delta_{1,0}^{\rm para}(\Omega)\Lambda_{1,0}^{\tau_1}(\omega; \Omega) G_{\bm k}(\omega)
]^<
\notag
\\
&\quad
+\frac{i}{6} \int \frac{d\omega}{2\pi} \frac{1}{N} \sum_{\bm k}
{\rm Tr}[
\dot\epsilon_{\bm k} G_{\bm k}(\omega+\Omega) \dot\epsilon_{\bm k} G_{\bm k}(\omega) \ddot\Delta_{0,0}^{\rm dia}(\Omega)\Lambda_{0,0}^{\tau_1}(\omega; \Omega) G_{\bm k}(\omega)
]^<
\notag
\\
&\quad
+\frac{i}{3} \int \frac{d\omega}{2\pi} \frac{1}{N} \sum_{\bm k}
{\rm Tr}[
\dot\epsilon_{\bm k} G_{\bm k}(\omega+\Omega) \dot\epsilon_{\bm k} G_{\bm k}(\omega+\Omega) \ddot\Delta_{1,0}^{\rm dia}(\Omega)\Lambda_{1,0}^{\tau_1}(\omega; \Omega) G_{\bm k}(\omega)
]^<
\notag
\\
&\quad
+\frac{i}{6} \int \frac{d\omega}{2\pi} \frac{1}{N} \sum_{\bm k}
{\rm Tr}[
\dot\epsilon_{\bm k} G_{\bm k}(\omega+\Omega) \ddot\Delta_{0,0}^{\rm dia}(\Omega)\Lambda_{0,0}^{\tau_1}(\omega+\Omega; \Omega) G_{\bm k}(\omega+\Omega) \dot\epsilon_{\bm k} G_{\bm k}(\omega)
]^<
\notag
\\
&\quad
+\frac{i}{3} \int \frac{d\omega}{2\pi} \frac{1}{N} \sum_{\bm k}
{\rm Tr}[
\dot\epsilon_{\bm k} G_{\bm k}(\omega+\Omega) \ddot\Delta_{1,0}^{\rm dia}(\Omega)\Lambda_{1,0}^{\tau_1}(\omega; \Omega) G_{\bm k}(\omega) \dot\epsilon_{\bm k} G_{\bm k}(\omega)
]^<,
\\
\chi_{\rm H3}^{(3)}(\Omega; \Omega, 0, 0)
&=
\frac{i}{6} \int \frac{d\omega}{2\pi} \frac{1}{N} \sum_{\bm k}
{\rm Tr}[
G_{\bm k}(\omega+\Omega) \dot\epsilon_{\bm k} G_{\bm k}(\omega) \ddot\Delta_{0,0}^{\rm para}(\Omega)\Lambda_{0,0}^{\tau_1}(\omega; \Omega) G_{\bm k}(\omega)
]^<
\notag
\\
&\quad
+\frac{i}{3} \int \frac{d\omega}{2\pi} \frac{1}{N} \sum_{\bm k}
{\rm Tr}[
G_{\bm k}(\omega+\Omega) \dot\epsilon_{\bm k} G_{\bm k}(\omega+\Omega) \ddot\Delta_{1,0}^{\rm para}(\Omega)\Lambda_{1,0}^{\tau_1}(\omega; \Omega) G_{\bm k}(\omega)
]^<
\notag
\\
&\quad
+\frac{i}{6} \int \frac{d\omega}{2\pi} \frac{1}{N} \sum_{\bm k}
{\rm Tr}[
G_{\bm k}(\omega+\Omega) \ddot\Delta_{0,0}^{\rm para}(\Omega)\Lambda_{0,0}^{\tau_1}(\omega+\Omega; \Omega) G_{\bm k}(\omega+\Omega) \dot\epsilon_{\bm k} G_{\bm k}(\omega)
]^<
\notag
\\
&\quad
+\frac{i}{3} \int \frac{d\omega}{2\pi} \frac{1}{N} \sum_{\bm k}
{\rm Tr}[
G_{\bm k}(\omega+\Omega) \ddot\Delta_{1,0}^{\rm para}(\Omega)\Lambda_{1,0}^{\tau_1}(\omega; \Omega) G_{\bm k}(\omega) \dot\epsilon_{\bm k} G_{\bm k}(\omega)
]^<.
\end{align}

\twocolumngrid

\section{Third harmonic generation}
\label{app: thg}

\begin{figure}[ht]
\begin{center}
\includegraphics[width=8cm]{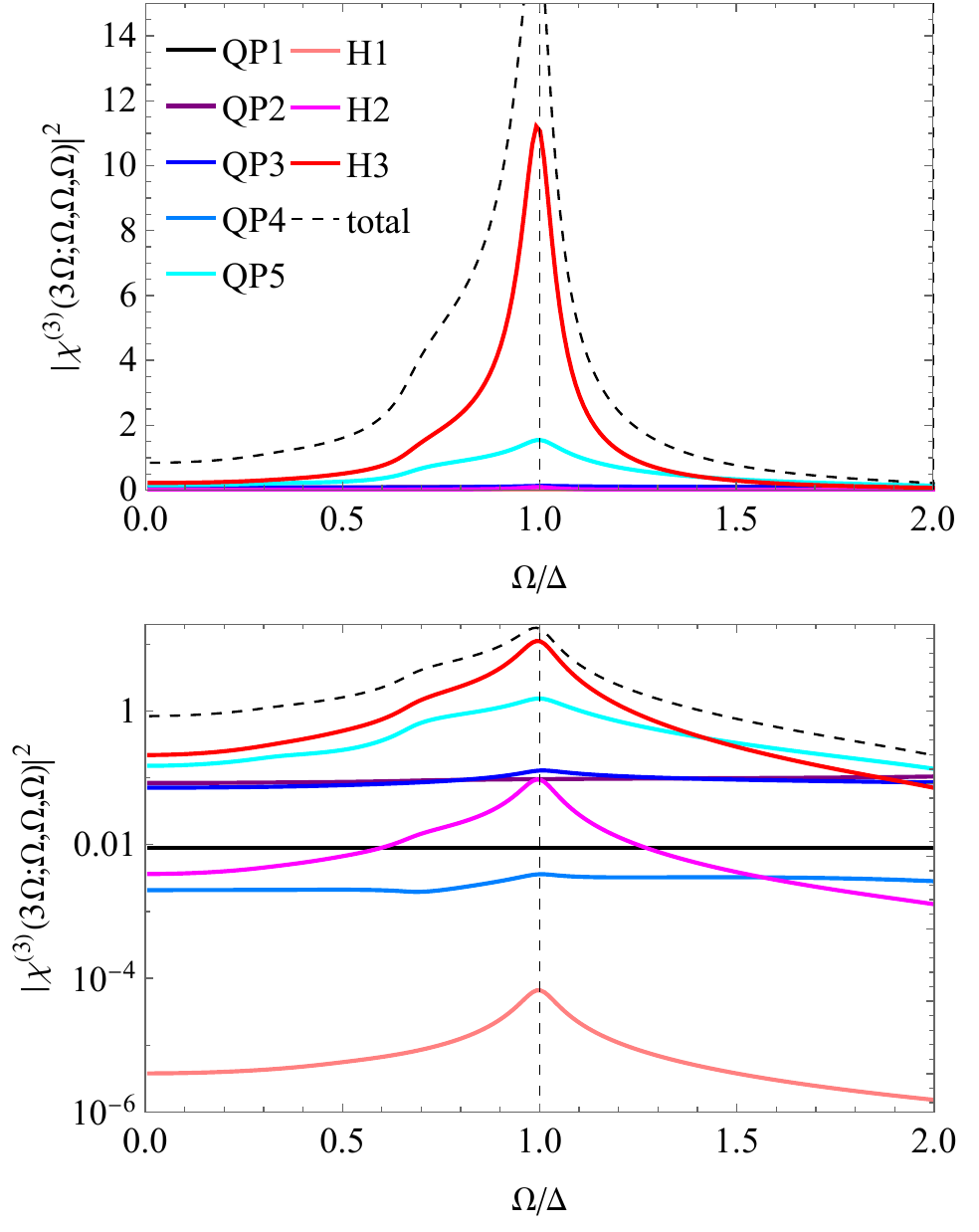}
\caption{Intensity of the THG susceptibility $\chi^{(3)}(3\Omega; \Omega, \Omega, \Omega)$
as a function of $\Omega/\Delta$ in the linear scale (top panel) and log scale (bottom) 
for the lattice model of disordered superconductors in the dirty regime.
The parameters are $V=2.5$, $\gamma=2$, and $\beta=50$. The vertical dashed
line indicates $\Omega=\Delta$.}
\label{fig: thg}
\end{center}
\end{figure}

In this section, we show the THG susceptibility $\chi^{(3)}(3\Omega; \Omega, \Omega, \Omega)$
for the single-band lattice model of disordered superconductors in the dirty regime. We follow the derivation of the THG susceptibility
detailed in Ref.~\cite{TsujiNomura2020}. The results are shown in Fig.~\ref{fig: thg} (the top (bottom) panel is plotted in the linear (log) scale), where one can find that
the THG susceptibility shows a resonance peak at $\Omega=\Delta$.
The dominant contribution comes from the H3 diagram, which is consistent with the previous work \cite{TsujiNomura2020}.
Contrary to the observation made in Ref.~\cite{Katsumi2024}, we do not find another peak centered around $\Omega=2\Delta/3$
and $\Omega=2\Delta$. There is, however, a little shoulder-like structure around $\Omega=2\Delta/3$, which has also been
observed in Ref.~\cite{Silaev2019}.

\bibliography{ref}

\begin{thebibliography}{79}
\bibitem [{rem()}]{remark}%
  \BibitemOpen
  \href@noop {} {}\bibinfo {note} {Typically, in lattice models one has $v_F
  \sim Wa$. In order to have a difference of several orders of magnitudes
  between $\frac{v_F}{W}$ and $L$, we need at least $L\gtrsim 10^2
  a$.}\BibitemShut {Stop}%
\bibitem [{pha()}]{phase_mode}%
  \BibitemOpen
  \href@noop {} {}\bibinfo {note} {In two dimensions, the phase mode can evade
  the Anderson-Higgs mechanism and remain to be a gapless mode. However, if we
  focus on spatially uniform ($q=0$) responses, the phase mode is decoupled
  from the electromagnetic response due to charge conservation and gauge
  invariance.}\BibitemShut {Stop}%
\end{thebibliography}

\end{document}